\documentclass[sigconf]{acmart}
\AtBeginDocument{%
  }

\setcopyright{acmlicensed}
\copyrightyear{2018}
\acmYear{2018}
\acmDOI{XXXXXXX.XXXXXXX}
\usepackage{multirow}
\usepackage{algorithm}      
\usepackage[most]{tcolorbox}
\usepackage{array}
\usepackage{tabularx} 
\usepackage{ragged2e} 
\usepackage{caption}  
\usepackage{algpseudocode} 
\usepackage[table]{xcolor}
\usepackage{float}
\setcopyright{none}
\acmConference[Conference acronym 'XX]{Make sure to enter the correct
  conference title from your rights confirmation email}{June 03--05,
  2018}{Woodstock, NY}
\acmISBN{978-1-4503-XXXX-X/2018/06}

\begin{document}

\title{When RAG Meets Query Planning: Logical Query Trees for Resolving Exploratory Reasoning Problems}

\author{Ganlin Xu}
\affiliation{%
 \department{School of Data Science}
  \institution{Fudan University}
  \city{Shanghai}
  \country{China}
}
\email{glxu24@m.fudan.edu.cn}

\author{Linghao Zhang}
\affiliation{%
 \department{School of Data Science}
  \institution{Fudan University}
  \city{Shanghai}
  \country{China}
}
\email{22307130274@m.fudan.edu.cn}

\author{Zhitao Yin}
\affiliation{%
 \department{School of Data Science}
  \institution{Fudan University}
  \city{Shanghai}
  \country{China}
}
\email{ztyin22@m.fudan.edu.cn}

\author{Hongda Xi}
\affiliation{%
 \department{School of Data Science}
  \institution{Fudan University}
  \city{Shanghai}
  \country{China}
}
\email{hdxi22@m.fudan.edu.cn}

\author{Chen Yang}
\affiliation{%
 \department{School of Data Science}
  \institution{Fudan University}
  \city{Shanghai}
  \country{China}
}
\email{yang_c25@m.fudan.edu.cn}

\author{Jiaqing Liang}
\affiliation{%
 \department{School of Data Science}
  \institution{Fudan University}
  \city{Shanghai}
  \country{China}
}
\email{liangjiaqing@fudan.edu.cn}

\author{Weijia Lu}
\affiliation{%
  \institution{United Automotive Electronic Systems}
  \city{Shanghai}
  \country{China}
}
\email{alfredwjlu@gmail.com}

\author{Sihang Jiang}
\affiliation{%
 \department{College of Computer Science and Artificial Intelligence}
  \institution{Fudan University}
  \city{Shanghai}
  \country{China}
}
\email{jiangsihang@fudan.edu.cn}

\author{Yanghua Xiao}
\affiliation{%
 \department{College of Computer Science and Artificial Intelligence}
  \institution{Fudan University}
  \city{Shanghai}
  \country{China}
}
\email{shawyh@fudan.edu.cn}

\author{Deqing Yang}
\authornote{Corresponding author.}
\affiliation{%
 \department{School of Data Science}
  \institution{Fudan University}
  \city{Shanghai}
  \country{China}
}
\email{yangdeqing@fudan.edu.cn}
\renewcommand{\shortauthors}{Trovato et al.}

\begin{abstract}
Retrieval-Augmented Generation (RAG) effectively grounds large language models (LLMs) in external knowledge but struggles with \textbf{exploratory reasoning problems (ERPs)} that are the complex queries involving high uncertainty and ambiguity. Resolving ERPs requires complex reasoning with unclear paths, tending to result in retrieval noise and error accumulation. Furthermore, the absence of an end-to-end planning mechanism makes it difficult to generate effective trajectories for ERPs. Motivated by database query planning, we introduce \emph{PlanRAG}, an RAG framework that models ERPs of natural language as \textbf{logical query trees (LQTs)}. However, translating ERPs into LQTs is non-trivial due to representation and optimization gaps between structured SQL and unstructured natural language, making it highly challenging to construct high-quality LQTs. To address these problems, we first decompose ERPs into atomic queries and then organize them into LQTs using dynamic programming guided by a cost model involving multiple complementary dimensions. Finally, we execute iterative aggregation, rewriting, retrieval, and generation over LQTs, processing nodes concurrently and propagating intermediate results upward, with further parallelization across multiple threads for efficiency. Our experimental results show that PlanRAG outperforms state-of-the-art iteration-based and graph-based RAG systems on our newly constructed dataset, \textbf{WikiWeb-ERP}, thereby providing a new formulation for optimizing natural language queries. Our source code and dataset are available at \url{https://anonymous.4open.science/r/PlanRAG-main-B2C8/}.
\end{abstract}


\received{20 February 2007}
\received[revised]{12 March 2009}
\received[accepted]{5 June 2009}

\maketitle

\section{Introduction}\label{sec:intro}
\begin{figure}
  \centering
  \includegraphics[width=\linewidth]{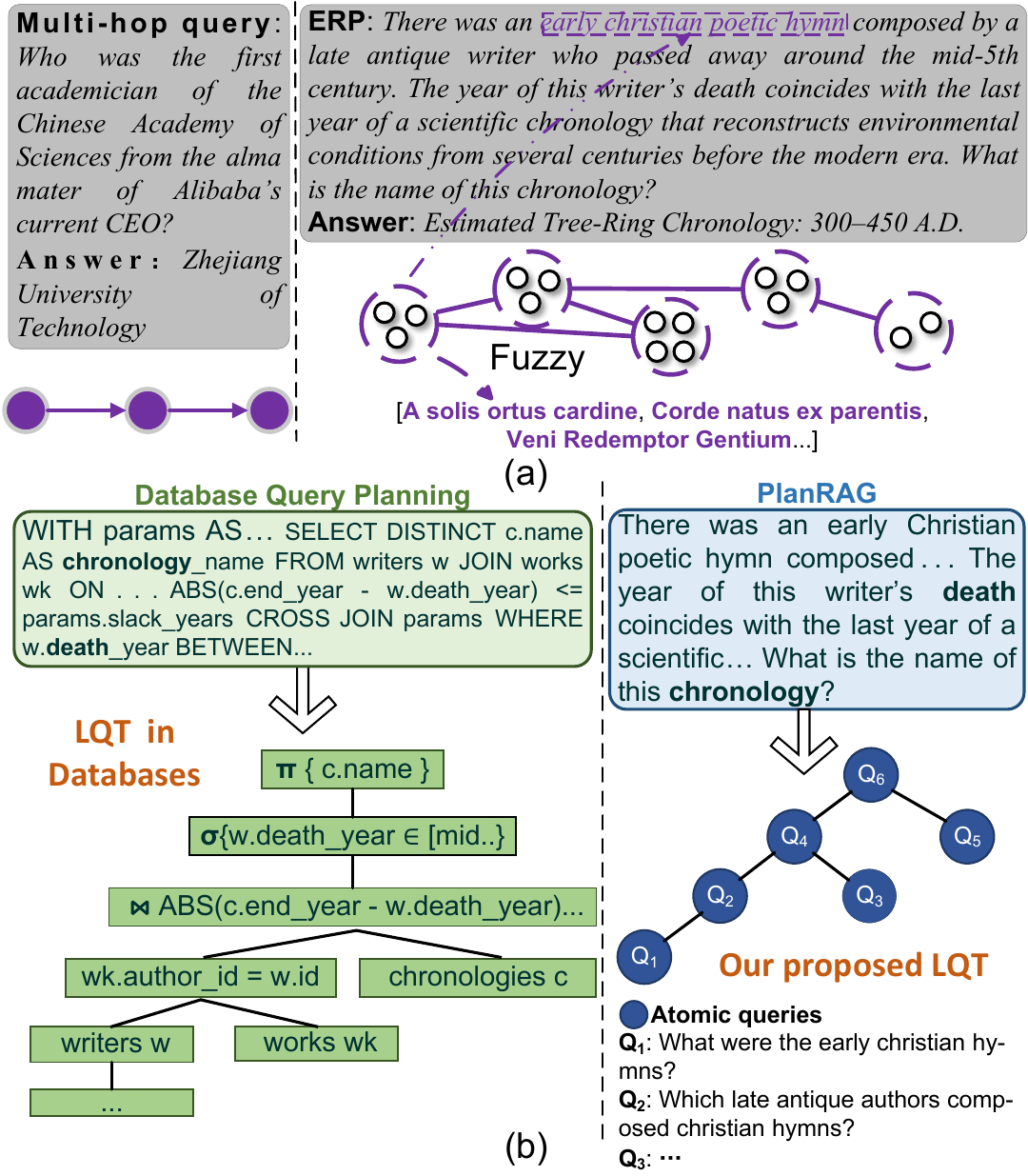}
  \caption{(a) A comparison between an ERP, where entities are coupled in fuzzy and emergent ways, and a conventional multi-hop query with explicit and decomposable reasoning chains.} (b) Motivated by database query planning, our PlanRAG framework models ERPs as logic query trees (LQTs).\label{figure1}
\end{figure}

Retrieval-Augmented Generation (RAG) integrates large language models (LLMs) with external document repositories, grounding responses from retrieved documents rather than parametric memory alone, thereby effectively mitigating LLMs' hallucinations~\cite{NEURIPS2020_6b493230, gao2024retrievalaugmentedgenerationlargelanguage}.  Owing to its ability to combine retrieval and generation, RAG has demonstrated strong performance on knowledge-intensive tasks that require factual grounding and evidence-based reasoning. However, existing RAG systems largely overlook the complex queries referred to as \textbf{exploratory reasoning problems (ERPs)}\footnote{The exploratory reasoning problems in this paper correspond to what is referred to as ``level 3 tasks'' in~\cite{mialon2023gaiabenchmarkgeneralai, li2025websailornavigatingsuperhumanreasoning}} in this paper, as shown in Figure~\ref{figure1}(a). ERPs are characterized by high uncertainty and difficulty in reduction, where entities are coupled in fuzzy and emergent ways ~\cite{mialon2023gaiabenchmarkgeneralai, li2025websailornavigatingsuperhumanreasoning}.

The inherent complexity of ERPs poses fundamental challenges for current RAG systems. As shown in Figure~\ref{figure1}(a), unlike conventional multi-hop queries with well-specified and decomposable chains, ERPs lack an explicit intermediate structure, making it unclear what information should be retrieved, in what order, and how to aggregate it coherently. More importantly, ERPs represent a broader class of scenarios that require long-horizon reasoning and global planning under uncertainty, in which sub-tasks are coupled by complex, implicit dependencies. Such characteristics make it inherently difficult to design effective planning strategies or decomposition schemes using existing methods, highlighting the fundamental importance of addressing ERPs. Consequently, existing RAG systems often struggle with the following three interrelated issues when resolving ERPs. ($i$) \textbf{Retrieval noise proliferates}: Due to ERPs' uncertainty and ambiguity, the retrieval process often returns large volumes of irrelevant documents, overwhelming LLMs with spurious evidence~\cite{li2025websailornavigatingsuperhumanreasoning}. ($ii$) \textbf{Error accumulation}: Iteration-based RAG methods process sub-queries sequentially, where early retrieval or generation errors propagate through later steps, leading to compounding hallucinations and unstable reasoning~\cite{gao2024retrievalaugmentedgenerationlargelanguage}. ($iii$) \textbf{Lack of end-to-end planning}: More importantly, existing planning approaches rely on manually annotated trajectories or decomposition rules, lacking an end-to-end mechanism for query optimization~\cite{li2025websailornavigatingsuperhumanreasoning, wu2025webdancerautonomousinformationseeking}. \emph{We observe that a \textbf{deterministic global planning mechanism} can not only coordinate the execution order and dependencies across sub-queries, but also effectively reduce retrieval noise and error accumulation}~\cite{li2025websailornavigatingsuperhumanreasoning}. Notably, the limitation, i.e., the lack of an end-to-end planning to coordinate retrieval and generation, structurally resembles the core challenge addressed by query planning in database systems, where the order and dependencies of operations must be optimized before execution. This analogy motivates us to explore whether a planning mechanism can benefit natural language queries by introducing structured paths and reducing ambiguity in retrieval.

As shown in Figure~\ref{figure1}(b), motivated by database query planning where SQL queries are compiled into \textbf{logical query trees (LQTs)} and optimized through cost-based strategies~\cite{10.1145/582095.582099}, we propose to construct LQTs from natural language queries of ERPs, enabling efficient and interpretable planning. However, translating this idea into natural language is non-trivial due to two fundamental gaps between structured SQL and unstructured natural language. ($i$) \textbf{Representation gap}: SQL queries are grounded in explicit relational schemas that predefine how relational atoms can be joined via well-specified keys and relations, providing a clear blueprint for LQT representation. In contrast, natural language queries are unstructured, where semantic units and latent dependencies among them remain implicit and ambiguous. There is no predefined schema to dictate how these units should be organized into LQTs, making the automatic derivation of structurally robust and semantically sound LQTs a core challenge. ($ii$) \textbf{Optimization gap}: Database optimizers evaluate candidate LQTs in SQL using cost models that estimate physical execution cost (I/O, CPU, and memory) based on well-defined schemas and statistics. In contrast, optimizing LQTs in natural language cannot rely on such physical metrics. Therefore, the cost must reflect novel perspectives, such as semantics and structure,  necessitating a multi-dimensional cost model tailored for conceptual query planning.

To bridge these gaps, we introduce an end-to-end RAG framework \emph{PlanRAG}, that models ERPs as LQTs for ERPs, thereby mitigating retrieval noise and error accumulation. The overall pipeline of PlanRAG consists of three main stages, which align with classical database query planning: \emph{query parsing}, \emph{logical optimization}, and \emph{physical execution}\footnote{Our current framework focuses primarily on the logical optimization, while physical optimizations such as caching
mechanisms are left for future work.}. First, in the query parsing, to address the \textbf{representation gap}, we decompose ERPs into atomic queries, each corresponding to a minimal semantic unit, i.e., a single relational triple, enabling precise representation of intermediate dependencies. These atomic queries exhibit semantic relationships, including \emph{Parent–child}, \emph{Sibling}, and \emph{Unrelated}. Second, in the logical optimization, we hierarchically organize these atomic queries into an optimal LQT via semantic relationships using Selinger's dynamic programming (DP). Before DP, we introduce relation preprocessing to reduce the number of LLM calls required to evaluate the potential node merging. During DP, we propose cycle prevention to preserve the DAG structure of LQTs, and context-aware merging to ensure that each local merging aligns with the tree structure. To address the \textbf{optimization gap} and guide the construction of high-quality LQTs, we design a cost model from five complementary dimensions, i.e., \emph{tree size} (node number), \emph{structural density} (edge number), \emph{tree depth}, \emph{tree balance} and \emph{semantic similarities}, ensuring that LQTs are structurally robust and semantically sound. Finally, in the physical execution, we execute aggregation, rewriting, retrieval, and generation in an iterative manner over LQTs, where nodes are processed concurrently and intermediate results are recursively propagated upward until the root node yields the final answer. To mitigate latency and enable efficient inference, we further parallelize execution by scheduling independent nodes concurrently using multi-threaded execution.

To evaluate PlanRAG's performance on ERPs, we have constructed and released an RAG dataset, namely \textbf{WikiWeb-ERP}, which contains some queries collected from existing datasets and those constructed following the approach of WebSailor~\cite{li2025websailornavigatingsuperhumanreasoning}, along with their corresponding documents retrieved from both Wikipedia and live webpages. Our experiments demonstrate our framework's superiority over some state-of-the-art baselines on WikiWeb-ERP, and its potential for handling ERPs. 

The main contributions of this paper include:

1. Following the paradigm of database query planning, we propose an RAG framework \emph{PlanRAG} for resolving ERPs, which improves the planning capability for complex retrieval tasks and provides a new paradigm for optimizing natural language queries.

2. To bridge the gaps between structured SQL and unstructured natural language queries, we propose an effective method to construct LQTs for ERPs via dynamic programming guided by a multi-dimensional cost model, which enables efficient and interpretable planning within the RAG paradigm and presents a new application scenario for database query optimization techniques.

3. To systematically evaluate our PlanRAG's performance specifically on ERPs, we further release a new dataset, WikiWeb-ERP, as a complex RAG benchmark. Our experimental results demonstrate that PlanRAG achieves state-of-the-art performance on WikiWeb-ERP over existing iteration-based and graph-based RAG systems, highlighting the critical advantage of our planning mechanism.

\section{Related Work}
\subsection{Retrieval-Augmented Generation}
Retrieval-Augmented Generation (RAG) integrates the generative capability of large language models (LLMs) with external knowledge retrieval, enabling dynamic access to external knowledge sources during reasoning~\cite{NEURIPS2020_6b493230, izacard2021distilling, ram-etal-2023-context}. This integration substantially enhances model performance on tasks requiring external knowledge retrieval and reasoning, such as question answering~\cite{li-etal-2025-raspberry}, fact verification~\cite{yue-etal-2024-retrieval}, and knowledge-grounded dialogue~\cite{cheng2025surveyknowledgeorientedretrievalaugmentedgeneration}. The core idea of RAG is to employ a retriever to identify documents or evidence passages relevant to a given query from large-scale corpora, and to condition the generator on the retrieved evidence to produce grounded responses. Consequently, RAG effectively alleviates the limitations of LLMs’ parameterized knowledge and extends their applicability to tasks requiring real-time information access, broad knowledge coverage, and cross-document reasoning~\cite{izacard-grave-2021-leveraging}.
\subsection{RAG for Multi-Hop QA}
Multi-hop question answering (QA) requires the model to perform a sequence of interdependent reasoning steps, thereby constructing a reasoning chain that leads to the final answer~\cite{ramesh-etal-2023-single, sagirova-burtsev-2023-uncertainty}. Multi-hop QA constitutes a compelling testbed for evaluating RAG systems because it imposes stringent requirements on both the knowledge retriever and the answer generator~\cite{zhuang-etal-2024-efficientrag, lee-etal-2025-magic}. Graph-based methods explicitly model multi-hop dependencies by constructing document/reasoning graphs and retrieving evidence along graph paths, thereby enhancing coherence and interpretability~\cite{liu-etal-2025-hoprag, Cao_2025}. Iterative-based RAG methods enable models to repeatedly access external knowledge during the reasoning process, progressively constructing coherent reasoning chains and thereby effectively addressing challenges in multi-hop inference~\cite{shao-etal-2023-enhancing, shi-etal-2024-generate}. Unlike iterative-based methods that execute sub-queries sequentially in a simple chain, ERPs examined in this paper involve problems with high uncertainty and difficulty in reduction \cite{wei2025browsecompsimplechallengingbenchmark}. ERPs cannot be adequately addressed through straightforward sequential reasoning; instead, they necessitate well-designed query planning that optimizes the execution order of sub-queries and the dependencies in the query structure.
\subsection{Information-Seeking Agents}
Information-seeking agents~\cite{li2025webthinkerempoweringlargereasoning, wu-etal-2025-webwalker, qiao2025webresearcherunleashingunboundedreasoning, geng2025webwatcherbreakingnewfrontier, li2025webweaverstructuringwebscaleevidence, wu2025resumunlockinglonghorizonsearch} aim to develop autonomous web systems capable of issuing search queries, navigating webpages, and extracting relevant evidence through sequential decision-making. Recent studies have demonstrated that these agents perform well on ERPs~\cite{li2025websailornavigatingsuperhumanreasoning, wu2025webdancerautonomousinformationseeking,tao2025webleaperempoweringefficiencyefficacy}, which involve complex, long-horizon information-seeking tasks. Such agents typically rely on large collections of high-quality interaction trajectories for supervision, leveraging reinforcement learning or imitation learning to enhance performance. Fundamentally, information-seeking agents perform online optimization over action-sequence search spaces, where planning and execution are tightly interleaved and decisions are refined incrementally based on intermediate observations. This online paradigm, while flexible, often incurs high training costs and substantial variance during inference. In contrast, PlanRAG adopts offline planning, where planning is optimized before execution. Rather than searching over action sequences, PlanRAG explores a tree-structured search space, i.e., LQTs, directly from the natural language query. This shift enables global structural optimization without relying on trajectory supervision or online policy learning. As a result, our method is training-free and does not require manually annotated trajectories, offering greater practicality and scalability across domains.
\begin{figure*}
  \centering
  \includegraphics[width=\linewidth]{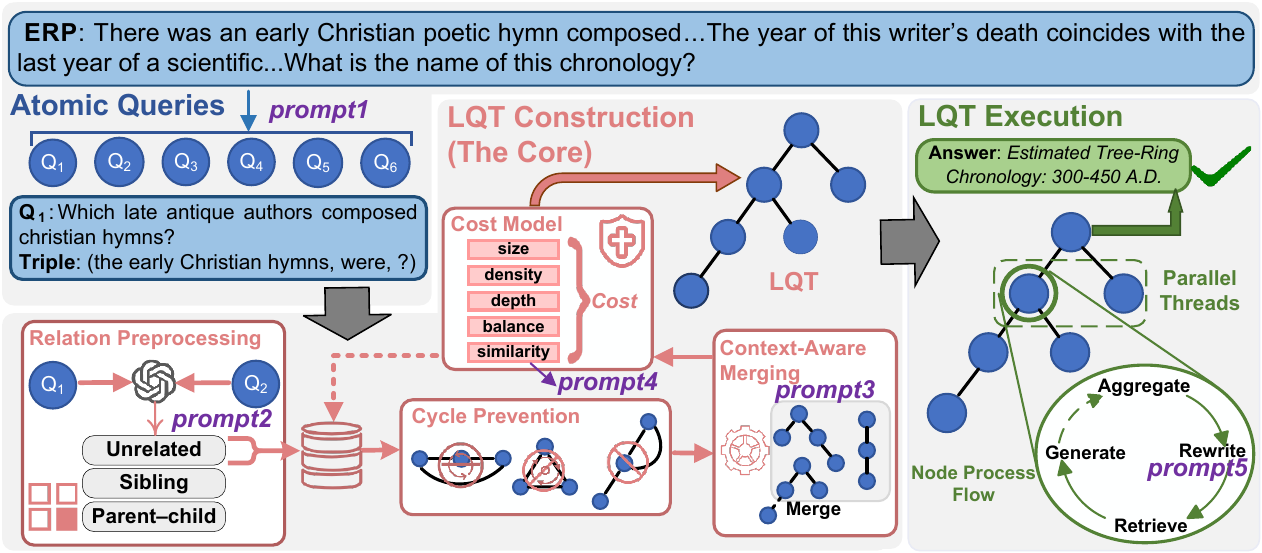}
  \vspace{-0.5cm}
  \caption{The pipeline of our proposed PlanRAG.
  }\label{figure2}
\end{figure*}

\section{Preliminaries}
\subsection{Query Planning}
In database systems, SQL queries are first parsed into an abstract syntax tree and then transformed into logical query trees (LQTs), which capture queries' semantics in relational algebra~\cite{10.1145/3626246.3654692}. Based on the logical representation, query optimizers systematically explore alternative query plans through equivalence-preserving transformations, such as operator reordering and decomposition. Candidate plans are evaluated using cost models that estimate execution efficiency, and the optimized logical plan is subsequently translated into a physical query plan for execution. This planning process enables databases to perform global reasoning about query execution before accessing actual data.

\begin{table}[t]
\centering
\caption{Atomic queries derived from the ERP in Figure~\ref{figure2}.}
\label{tab1}
\vspace{-0.2cm}
\begin{tabular}{p{0.03\linewidth} p{0.87\linewidth}}
\toprule
\textbf{Q$_i$} & \textbf{Atomic Queries} \\
\midrule
Q$_1$ & What were the early Christian hymns? \\
Q$_2$ & Which late antique authors composed Christian hymns? \\
Q$_3$ & Which authors passed away around the middle of the 5th century? \\
Q$_4$ & Which authors’ years of death coincide with the final year of a scientific chronology? \\
Q$_5$ & Which scientific chronologies reconstruct environmental conditions from several centuries before the modern era? \\
Q$_6$ & What is the name of the scientific chronology? \\
\bottomrule
\end{tabular}
\vspace{-0.2cm}
\end{table}

\subsection{Logical Query Tree}
As shown on the left in Figure~\ref{figure1}(b), a logical query tree (LQT) is typically defined as a hierarchical logical plan derived from structured queries (e.g., SQL) in database systems, where nodes correspond to relational algebra operators (such as selection, projection, and join) and edges represent data flow between operators~\cite{10.1145/3183713.3190655}. These trees are grounded in well-defined schemas and equivalence rules, ensuring deterministic semantics and enabling cost-based optimization for efficient query execution.

In this paper, we redefine LQTs as hierarchical representations that capture the compositional semantics of natural language queries. Formally, an LQT is modeled as a directed acyclic graph (DAG) $\mathcal{T} = (V, E)$, where $V$ is the node set and $E$ is the edge set. Each node $v \in V$ represents the atomic queries, as shown in Table~\ref{tab1}, which are derived from the original ERP. Each edge $e = (v_i, v_j) \in E$ encodes semantic dependencies between atomic queries, with leaf nodes corresponding to directly executable atomic queries. The LQT execution follows a bottom-up process: nodes are processed concurrently and intermediate results are recursively propagated upward until the root node yields the final answer. Unlike LQTs in databases, which operate over relations and algebraic operators with rigid semantics, LQTs in this paper are semantic composition trees constructed from natural language queries. This adaptation allows our LQTs to capture the compositional structure inherent in ERPs, providing a flexible yet interpretable system for RAG.

\subsection{Selinger’s Dynamic Programming}
Selinger’s dynamic programming (DP) is a classical method in database query planning for generating optimized LQTs from a set of atomic queries~\cite{10.1145/275487.275492, 10.1145/582095.582099}. The input of the algorithm is a set of atomic queries, each representing an indivisible basic operation derived from the original query, and the output is an optimized LQT \(\mathcal{T} = (V, E)\), where the nodes \(V\) correspond to atomic or intermediate sub-queries and the edges \(E\) encode semantic dependencies between sub-queries. The algorithm follows a bottom-up strategy: it first computes the execution cost for each atomic query and then recursively combines smaller subplans into larger ones for progressively larger subsets of atomic queries, selecting the combination with the minimum cost at each step. Let the set of atomic queries be \(\mathcal{Q} = \{q_1, q_2, \ldots, q_n\}\). For any non-empty subset \(S \subseteq \mathcal{R}\), let \(C(S)\) denote the minimum estimated cost for executing all joins among the queries in \(S\). To compute cost \(C(S)\), the algorithm considers splitting \(S\) into two non-empty subsets, where \(A \subset S\) represents one subset and the other subset is \(S \setminus A\). The recursive cost formula is defined as:
\begin{equation}
C(S)
= \min_{A \subset S}
\left\{
C(A) + C(S \setminus A) + J(A, S \setminus A)
\;\middle|\;
A \neq \emptyset
\right\},
\end{equation}
where \(J(A, S \setminus A)\) estimates the incremental cost of joining the subplan over \(A\) with the subplan over \(S \setminus A\), including I/O, CPU, and memory usage. The algorithm enumerates subsets in increasing order of size and prunes suboptimal intermediate plans to ensure that each intermediate subplan is locally optimal. Ultimately, all locally optimal subplans are combined to form a globally optimal LQT \(\mathcal{T}\). Compared with greedy or heuristic tree generation, DP supports systematic pruning, stable cost-based evaluation, and natural integration of semantic signals without relying on gradient-based learning. These properties make Selinger's DP particularly suitable for LQT construction under semantic uncertainty in natural language. 
\subsection{BGE Model}
BGE~\cite{10.1145/3626772.3657878} is a pretrained dense embedding model designed for semantic similarity estimation and retrieval tasks. It maps input text into a dense vector space, where semantically similar texts are close to each other. Formally, given two text sequences $x$ and $y$, BGE encodes them into dense representations via an encoder $E(\cdot)$, and their 
similarity is computed as:
\begin{equation}
\text{sim}(x, y) = \langle E(x), E(y) \rangle,
\end{equation}
where $\langle \cdot, \cdot \rangle$ denotes the inner product. In this paper, we use BGE to encode both the original query and the textualized logical query tree. The resulting 
similarity score measures the semantic alignment between the constructed logical query tree and the original query, and serves as a key signal in our cost model (Section~\ref{LogicalQueryTreeConstruction}).
\section{Methodology}\label{method}
\subsection{Overview}
The pipeline of our proposed PlanRAG is shown in Figure~\ref{figure2}, which enables global planning for ERPs. We begin with decomposing each ERP into atomic queries, namely, minimal semantic units (atomic query generation). Analogous to LQT construction in database query planning, we then organize these atomic queries into optimal LQTs via dynamic programming guided by a multi-dimensional cost model (LQT construction). To ensure efficiency, we perform a bottom-up and parallel execution strategy over LQTs, involving iterative aggregation, rewriting, retrieval, and generation until the root node yields the final results (LQT execution). 

Notably, corresponding to the database query processing pipeline, the three stages of our PlanRAG are query parsing (atomic query generation), logical optimization (LQT construction), and physical execution (LQT execution). Our approach focuses primarily on the logical optimization, while physical optimization (e.g., caching mechanisms) is reserved for future research.

\subsection{Atomic Query Generation}
To enable precise representation of intermediate dependencies, we employ well-designed prompting techniques to leverage LLMs for decomposing complex queries into multiple atomic queries, where the detailed prompts are introduced in Appendix A. Specifically, we provide an explicit definition of atomic queries together with illustrative demonstrations in the prompt, guiding the LLM to formulate each atomic query in the form of (subject, predicate, object).  In this paper, each atomic query is indivisible and corresponds to a single relational triple. For example, the atomic query (``What were the early Christian hymns?'') derived from the ERP in Figure~\ref{figure2} corresponds to a triple \emph{(the early Christian hymns, were, ?)}. This strict one-to-one mapping ensures both the semantic atomicity of decomposition results and the natural mapping between atomic queries and triples, thereby preserving the logical structure of original queries and enabling precise retrieval of supporting documents. These atomic queries can be organized into LQTs based on three types of semantic relationships: ($i$) \textbf{Parent–child:} One atomic query serves as the parent, while the other serves as the child, with the child semantically depending on or refining the parent. ($ii$) \textbf{Sibling:} Two queries are semantically independent and serve as sibling nodes at the same level of LQTs. ($iii$) \textbf{Unrelated:} Two queries have no meaningful semantic connection and do not share any parent–child or sibling relationship in LQTs.

\subsection{Logical Query Tree Construction}\label{LogicalQueryTreeConstruction}
Analogous to LQT construction in database query planning, we extend Selinger's dynamic programming (DP) paradigm~\cite{10.1145/582095.582099} to organize these atomic queries into LQTs. Before DP, we incorporate \textbf{relation preprocessing} to reduce the number of LLM calls. During DP, we propose \textbf{cycle prevention} to ensure that LQTs viewed as directed acyclic graphs (DAGs) stay cycle-free, and \textbf{context-aware merging} to preserve global semantic coherence, ensuring that each local merging aligns with the tree structure. Throughout this process, we evaluate all candidate trees using a multi-dimensional \textbf{cost model} to optimize LQTs. The pseudocode for LQT construction procedure is listed in Appendix B.

\paragraph{Relation Preprocessing} 
To reduce the number of LLM calls required to evaluate potential node merging during DP, we introduce relation preprocessing before DP. In our framework, unrelated relationships between atomic queries are determined during a relation preprocessing stage rather than during the DP process. Given an ERP, we utilize LLMs to semantically evaluate all pairs of atomic queries $(q_i, q_j)$ and categorize them into three types of relationships (`Unrelated', `Sibling' and `Parent-child'). Specifically, the notion of unrelated query pairs, together with representative examples, is incorporated into the LLM prompt to guide the model in identifying unrelated atomic query pairs. The identified relationships are cached during the stage. As shown in Figure~\ref{figure2}, we mark unrelated relationships between atomic queries. By observing a large number of samples, we find that LLMs can accurately identify unrelated query pairs (Section~\ref{fine-grainedanalysisonlqtconstruction}). Therefore, all query pairs labeled as ``Unrelated'' are immediately pruned without LLM invocation during DP. The detailed prompts for determining relationships between atomic queries are introduced in Appendix A. 

\paragraph{Cycle Prevention} 
To preserve the DAG structure of LQTs, the algorithm prevents cycles by first checking whether merging the two nodes would introduce a cycle during DP. If so, the merging is disallowed. 

\paragraph{Context-Aware Merging}
To ensure that each local node merging aligns with the tree structure, we introduce context-aware merging to determine whether the ``parent–child'' relationship exists between two nodes. The key idea is to incorporate the partially constructed LQT into the LLM prompt, enabling the evaluation of the semantic compatibility of candidate nodes. Consequently, the technique improves the quality of resulting LQTs by grounding each node merging in the evolving context. 

\paragraph{Cost Model} 
Unlike traditional database cost models that estimate physical execution cost (I/O, CPU, and memory) based on well-defined schemas and statistics, our proposed cost model is designed to evaluate the quality of LQTs from both structural and semantic perspectives. It evaluates LQTs in terms of structural soundness and derivational quality, while incorporating semantic consistency and alignment with the original query as optimization objectives. Specifically, the cost of a candidate tree is evaluated from the following five aspects\footnote{While the cost function is manually designed, each dimension (e.g., tree size, structural density) is chosen based on an intuitive understanding of LQT and error propagation, serving as an interpretable structural inductive bias to guide LQT construction toward robust structures.}: 

\noindent$\bullet$  \textbf{Tree size (ts)}: It is the number of nodes in LQTs. A larger tree size encourages the inclusion of as many atomic queries as possible, thereby preventing the loss of essential semantic units and ensuring that the reasoning chain remains complete.

\noindent$\bullet$  \textbf{Structural density (sd)}: It is the number of edges in LQTs. A sufficiently dense structure reflects rich semantic dependencies among atomic queries, which helps retain their interconnectedness.

\noindent$\bullet$  \textbf{Tree depth (td)}: It is the maximum distance from the root to any leaf node. Controlling depth prevents trees from degenerating into a long linear chain, which introduces sequential bottlenecks and amplifies propagation.

\noindent$\bullet$  \textbf{Tree balance (tb)}: The tree’s balance is measured by subtree height differences. A balanced tree structure helps further improve parallel efficiency while preventing the tree from degenerating into an overly deep linear chain.

\noindent$\bullet$  \textbf{Semantic similarity (ss)}: We transform LQTs into natural language and measure the similarity to the original query using BGE~\cite{10.1145/3626772.3657878}. This dimension guides the optimizer by enforcing alignment with the original query intent, anchoring the constructed tree to the original query and ensuring that the planning remains semantically accurate. The full prompts are introduced in Appendix A.

Formally, let the candidate tree be $\mathcal{T} = (V, E)$, where $V$ and $E$ denote the set of nodes (atomic queries) and directed edges, respectively. The total cost is defined as:
\begin{equation}
\text{Cost}(\mathcal{T}) =
- \Bigl( |V| + |E| - \text{Depth}(\mathcal{T})
- \text{Balance}(\mathcal{T})
+ \alpha \, \text{Sim}(\mathcal{T}) \Bigr),
\label{eq2}
\end{equation}
where $|V|$ denotes $\mathcal{T}$'s size, $|E|$ represents $\mathcal{T}$'s structural density, $\text{Depth}(\mathcal{T})$ measures $\mathcal{T}$'s depth, $\text{Balance}(\mathcal{T})$ quantifies $\mathcal{T}$'s balance, and $\text{Sim}(\mathcal{T})$ captures the semantic similarity. Tree size $|V|$ and structural density $|E|$ are treated as positive contributors, encouraging coverage of more semantic units. In contrast, tree depth $\text{Depth}(\mathcal{T})$ and tree balance $\text{Balance}(\mathcal{T})$ are included as negative terms to discourage overly deep or unbalanced structures. The semantic similarity term $\text{Sim}(\mathcal{T})$ is scaled by a factor of $\alpha$ ($\alpha > 1$) to amplify its role in aligning the constructed tree with the original query intent, thereby avoiding semantic deviation. The cost model is a multi-objective trade-off, not a simple preference for larger plans. Tree size and structural density serve as positive terms, while tree balance and depth act as negative terms. This design encourages high coverage of atomic queries while preventing the plan from degenerating into linear structures. In addition to this, redundancy is discouraged by the semantic similarity term (weighted by $\alpha$), which measures each node’s alignment with the original query intent. Repetitive or irrelevant substructures reduce this score and are penalized. Furthermore, the DP search evaluates alternative structures covering the same atomic queries, selecting the plan with the lowest cost. Redundant decompositions often increase depth or reduce semantic alignment, making them unlikely choices. To simplify the formulation and maintain balanced contributions across dimensions during initial optimization, we assign equal linear coefficients to each dimension except for $\text{Sim}(\mathcal{T})$. The DP algorithm enumerates all possible construction orders and selects the optimal LQT with the minimal cost:
\begin{equation}
\mathcal{T}^* = \arg\min_{\mathcal{T}} \text{Cost}(\mathcal{T}).
\end{equation}
Overall, unlike database query optimization—which relies on well-defined statistics and physical cost metrics—the ``optimal LQT'' in our approach is optimal only under a heuristic and LLM-driven function. Therefore, our method is better viewed as a structured search strategy rather than traditional cost-based optimization. Due to the heuristic nature of the cost estimates, the DP optimality guarantee from database query optimization does not directly apply to our LQT setting. Nevertheless, PlanRAG remains a structured and database-inspired planning method that effectively guides semantic compositions, without claiming equivalent theoretical guarantees.
\subsection{Execution Strategy over Logical Query Trees}
After obtaining the optimized LQT, we iteratively perform bottom-up aggregation, rewriting, retrieval, and generation over LQTs. In LQTs, each node $v$ represents an atomic query. The processing differs between leaf and non-leaf nodes.\\
\noindent \texttt{Leaf nodes} retrieve directly from the corpus and generate answers:
\begin{equation}
\mathrm{Ans}(v)
= \mathrm{Generate}\bigl(\mathrm{Retrieve}(v)\bigr).
\end{equation}
\noindent \texttt{Non-leaf nodes} first aggregate the queries and answers of their child nodes as context to guide rewriting the original queries. Then, we use the rewritten queries to retrieve passages and generate answers:
\begin{equation}
\begin{gathered}
\mathrm{Ctx}(v)
= \mathrm{Aggregate}\!\left(
\{ (v, u, \mathrm{Ans}(u)) \mid u \in \mathrm{Children}(v) \}
\right), \\
v'
= \mathrm{Rewrite}\!\left(v, \mathrm{Ctx}(v)\right), \\
\mathrm{Ans}(v')
= \mathrm{Generate}\!\left(
\mathrm{Retrieve}(v')
\right),
\end{gathered}
\end{equation}
where $\mathrm{Aggregate}(\cdot)$, $\mathrm{Rewrite}(\cdot)$, $\mathrm{Retrieve}(\cdot)$, and $\mathrm{Generate}(\cdot)$ are aggregation, rewriting, retrieval, and generation operations, respectively. The prompts for rewriting queries are introduced in Appendix A. To enhance computational efficiency and fully leverage parallelism, we follow the DAG's topological order and execute in parallel all nodes using multi-threading. This layer-by-layer execution continues until the root node is processed.
\subsection{Prompt Interaction and Orchestration}
Our framework involves multiple LLM prompts across different stages, which interact in a structured pipeline. Specifically, the output of each stage serves as the input to subsequent steps, forming a coherent information flow.\\
\indent (1) In the atomic query generation, the LLM decomposes the ERP into atomic queries using prompt1 in Figure~\ref{figure2}.\\
\indent (2) In the relation preprocessing stage, prompt2 in Figure~\ref{figure2} is applied to determine semantic relationships between atomic queries, and the identified unrelated pairs are cached for subsequent pruning during LQT construction. \\
\indent(3) During LQT construction, the LLM is further invoked for context-aware merging, where the partially constructed tree is incorporated into the prompt (prompt3 in Figure~\ref{figure2}) to guide local decisions.\\
\indent (4) In the cost modeling stage, we prompt (prompt4 in Figure~\ref{figure2}) the LLM to transform candidate LQTs into natural language queries, which are then compared with the original query to estimate semantic alignment.\\
\indent (5) In the execution stage, we use LLMs (prompt5 in Figure~\ref{figure5}) to aggregate the queries and answers of child nodes as context, and use this context to rewrite the atomic query for subsequent retrieval and answer generation.\\
\indent This staged design ensures that intermediate representations (atomic queries, relations, and partial LQTs) are explicitly passed between prompts, enabling coordinated steps across the pipeline.
\subsection{Complexity Analysis}
\paragraph{Time Complexity of LQT Construction}
Our method focuses on constructing LQTs using Selinger's dynamic programming algorithm. Although the algorithm has exponential theoretical time complexity, with a classical upper bound of $\mathcal{O}(n^{2} 2^{n})$, a single query is rarely decomposed into more than 15 atomic queries in practical ERP scenarios by our statistics. The cases with more than 15 atomic queries are extremely rare, whereas most queries involve fewer atomic queries, making the search space acceptable. Our approach remains feasible even in more complex scenarios involving over 20 atomic queries, while preserving plan quality and execution efficiency.  The dominant cost in ERPs does not come from planning but from LLM invocations. Redundant planning steps may introduce redundant retrieval and unnecessary LLM calls during planning. In contrast, the search times of DP are negligible compared with downstream costs. Therefore, identifying a better execution structure during planning provides practical benefits: it reduces unnecessary LLM calls, improves retrieval efficiency, and controls overall system cost. During DP, any pair labeled as `Unrelated' is immediately pruned without invoking the LLM. Let \( p_0 \) denote the accuracy of detecting `Unrelated' pairs, then the expected number of LLM calls during DP can be approximated as:
\begin{equation}
\mathbb{E}\!\left[N_{\mathrm{DP}}\right]
\;\simeq\;
(1 - p_0)\,\binom{n}{2}.
\end{equation}
\begin{tcolorbox}[
  colback=gray!15, 
  colframe=gray!50, 
  boxrule=0pt, 
  arc=2pt, 
  left=1pt, right=1pt, top=1pt, bottom=1pt,
  breakable, enhanced
]
\paragraph{Proof.}
We clarify that our claim refers to the complexity of LLM invocations rather than the DP enumeration itself. In classical Selinger-style dynamic programming (DP), the search space consists of all subsets of $Q$, leading to $O(2^n)$ candidate states. However, this exponential subset enumeration does not directly correspond to the number of LLM invocations in our framework.

\textbf{Assumption 1 (Pairwise factorization of transitions).}
We assume that the feasibility of extending a partial solution $S \subseteq \mathcal{Q}$ with an element $q_i \notin S$ can be determined by pairwise semantic compatibility:
\[
S \cup \{q_i\}
\models
\bigwedge_{q_j \in S} R(q_i, q_j),
\]
where $R(q_i,q_j)$ denotes a semantic relation evaluated by an LLM during preprocessing. This pairwise factorization is standard in relation-based decomposition settings and allows higher-order subset dependencies to be reduced to pairwise interactions.

Under this formulation, each unordered pair $(q_i, q_j)$ is evaluated at most once during relation preprocessing and cached for subsequent DP usage. Let:
\[
X_{ij}=\mathbb{I}\!\left[R(q_i,q_j)\right].
\]

We assume that a pair is not evaluated if it is classified as ``Unrelated'', and define $p_0$ as the probability of pruning such pairs. Then:
\[
\mathbb{P}(X_{ij} = 1) = 1 - p_0.
\]

Since there are $\binom{n}{2}$ possible pairs, and LLM invocations correspond exactly to evaluated pairs, we have:
\[
N_{\mathrm{LLM}} = \sum_{i<j} X_{ij}.
\]

Taking expectation and applying linearity:
\[
\mathbb{E}[N_{\mathrm{LLM}}]
= \sum_{i<j} \mathbb{E}[X_{ij}]
= (1 - p_0)\binom{n}{2}
= O(n^2).
\]

Therefore, the expected number of LLM invocations is quadratic in the number of atomic queries.
\paragraph{Discussion.}
Compared with the original DP formulation that explores an exponential number of subset combinations $O(2^n)$, relation preprocessing effectively transforms subset-level reasoning into a sparse pairwise relation graph. This reduces LLM usage from exponential-scale DP exploration to quadratic pairwise evaluation in expectation, i.e., $O(n^2)$.
\end{tcolorbox}
\noindent Without relation preprocessing, DP must repeatedly invoke the LLM to assess every candidate pair during search, causing the number of LLM calls to grow exponentially with $n$ and resulting in more several-fold LLM calls. In typical scenarios, the DP enumeration accounts for only a small fraction of the total runtime (around $5\%$), while retrieval and LLM calls dominate (about $90\%$). Thus, despite the exponential theoretical complexity of DP, relation preprocessing and a relatively small number of atomic queries keep the actual runtime affordable, demonstrating the practical efficiency of our methods in real-world applications. 

\paragraph{Time Complexity of LQT Execution}
The LQT execution follows a bottom-up dependency order, where each node can be processed only after all of its child nodes are processed. Independent nodes can be executed in parallel to reduce runtime, and the total execution cost is
\begin{equation}
T_\mathrm{total} = \mathcal{O}\!\bigl(d (T_{\mathrm{LLM}} + T_{\mathrm{retrieval}})\bigr),
\end{equation}
where $d$ is the LQT's depth, $T_{\text{LLM}}$ and $T_{\text{retrieval}}$ denote the average latency of each LLM call and retrieval, respectively. In the best case, where the LQT is well-balanced, the depth is $d = \log n$. The total parallel execution cost is
\begin{equation}
T_\mathrm{best} = \mathcal{O}\!\bigl(
\log n  (\,T_{\mathrm{LLM}} + T_{\mathrm{retrieval}}\,)
\bigr).
\end{equation}
In the worst case, the LQT degenerates into a linear chain of depth 
$n$, where all nodes are sequentially dependent, i.e., $d = n$. In this scenario, the execution cost becomes
\begin{equation}
T_\mathrm{worst} = \mathcal{O}\!\bigl(n (T_{\mathrm{LLM}} + T_{\mathrm{retrieval}})\bigr).
\end{equation}

\begin{table}[b]
\centering
\caption{Proportions of ERPs across various datasets. BC, BC-Plus and BC-LC are the abbreviations for BrowseComp~\cite{wei2025browsecompsimplechallengingbenchmark}, BrowseComp-Plus~\cite{chen2025browsecompplusfairtransparentevaluation}, and BrowseComp Long Context~\cite{wei2025browsecompsimplechallengingbenchmark}, respectively.}
\label{tab2}
\vspace{-0.2cm}
\begin{tabular}{@{}lcccc@{}}
\toprule
\textbf{Dataset} & \textbf{BC} & \textbf{BC-Plus} & \textbf{BC-LC} & \textbf{GAIA} \\
\midrule
\textbf{Proportions (\%)}& 23.8 & 23.6 &29.2 &16.1 \\
\bottomrule
\end{tabular}
\end{table}

\subsection{WikiWeb-ERP}
To the best of our knowledge, there is currently no dedicated benchmark specifically designed for ERPs. As shown in Table~\ref{tab2}, existing datasets contain a relatively low proportion of ERPs. To comprehensively evaluate our PlanRAG's performance specifically on ERPs, we have constructed and released a high-quality dataset \textbf{WikiWeb-ERP} tailored for RAG systems. The construction procedure of WikiWeb-ERP is as follows.

First, we collect ERPs from two complementary aspects to ensure diversity and representativeness. Specifically, ($i$) \textbf{Dataset Sampling}: We selected QA pairs from several existing datasets (BrowseComp~\cite{wei2025browsecompsimplechallengingbenchmark}, BrowseComp-Plus~\cite{chen2025browsecompplusfairtransparentevaluation}, BrowseComp Long Context~\cite{wei2025browsecompsimplechallengingbenchmark}, and GAIA~\cite{mialon2023gaiabenchmarkgeneralai}) to guarantee data reliability and cross-dataset comparability. ($ii$) \textbf{Generative Expansion}: We adopt the query-generation approach proposed by WebSailor~\cite{li2025websailornavigatingsuperhumanreasoning} to further expand these QA pairs, which models real-world Internet information flows in practical and open-domain scenarios.

Similarly, we construct the document collection from two sources. First, we combine dense and sparse retrieval techniques to retrieve candidate documents from the Wikipedia dump related to each query as the first document source. Second, we use the web collection method in WebSalior~\cite{li2025websailornavigatingsuperhumanreasoning} to crawl real webpages from the Internet for each query, forming the second document source. These webpages capture the distribution of real-world knowledge and inherent noise in open-domain information. Together, the two types of documents constitute the document collection of WikiWeb-ERP, ensuring authenticity and diversity of retrieval sources.\\
\indent Overall, WikiWeb-ERP comprises 3,536 queries and 53,682 documents. Of these queries, 2,405 originate from existing datasets, while 1,131 are generated via WebSailor~\cite{li2025websailornavigatingsuperhumanreasoning}. In terms of documents, 24,050 are retrieved from the Wikipedia dump, and 29,632 are crawled from real webpages. The average lengths are 93.48 words per query and 121.09 words per document, respectively.

\section{Experiments}\label{sec:exp}
\subsection{Evaluation Metrics}
Following prior work on RAG and multi-hop QA~\cite{cheng-etal-2025-dualrag}, we evaluate model performance using four complementary metrics: \textbf{Accuracy (Acc)}, \textbf{Exact Match (EM)}, \textbf{Token-level F1 (F1)} and \textbf{Semantic Accuracy (Acc$\dagger$)}. These metrics jointly assess factual correctness, lexical overlap, and semantic consistency, which are particularly important for ERPs.  ($i$) \textbf{Acc} measures whether the generated answer adequately captures the key content of ground-truth answers. ($ii$) \textbf{EM} measures the degree of exact match between generated and ground-truth answers. ($iii$) \textbf{F1} evaluates the token-level similarity between generated and ground-truth answers. ($iv$) We also compute \textbf{Acc$\dagger$} using gpt-3.5-turbo-instruct for a more thorough evaluation, where the LLM evaluates the correctness of generated answers, taking ground-truth answers as reference.

\subsection{Baselines}
To ensure a comprehensive evaluation, we compare our PlanRAG with DirectLLM and NaiveRAG~\cite{NEURIPS2020_6b493230}, along with two major families of advanced RAG systems. Iteration-based methods include RetGen~\cite{shao-etal-2023-enhancing}, GenGround~\cite{shi-etal-2024-generate}, DualRAG~\cite{cheng-etal-2025-dualrag} and KiRAG~\cite{fang-etal-2025-kirag}, which perform retrieval and generation iteratively along a sequential reasoning chain. In contrast, graph-based methods include ChainRAG~\cite{zhu-etal-2025-mitigating}, HopRAG~\cite{liu-etal-2025-hoprag}, and LEGO-GraphRAG~\cite{Cao_2025}, which construct document/reasoning graphs to capture multi-hop dependencies and retrieve evidence along graph paths.

\noindent$\bullet$  \textbf{DirectLLM} generates answers without retrieving external knowledge.

\noindent$\bullet$  \textbf{NaiveRAG} retrieves documents once and generates answers based on the retrieved documents.

\noindent$\bullet$  \textbf{RetGen} enhances LLMs with an iterative retrieval-generation synergy strategy to answer a multi-hop question. 

\noindent$\bullet$  \textbf{GenGround} iteratively generates and grounds answers with retrieved evidence to improve multi-hop question answering.

\noindent$\bullet$  \textbf{DualRAG} jointly ranks queries and documents using multi-view relevance scoring.

\noindent$\bullet$  \textbf{KiRAG} enhances iterative retrieval-augmented generation by using knowledge triples to build reasoning chains.

\noindent$\bullet$  \textbf{ChainRAG} iteratively rewrites sub-questions and retrieves sentences via a sentence graph to address missing entities for multi-hop QA.

\noindent$\bullet$  \textbf{HopRAG} constructs a passage graph using pseudo-queries and performs multi-hop reasoning through a `retrieve-reason-prune' mechanism.

\noindent$\bullet$\textbf{LEGO-GraphRAG} modularizes GraphRAG retrieval into subgraph-extraction and path-retrieval modules to enable flexible construction and systematic evaluation of instances.

\subsection{Implementation Details}
We conduct all experiments on three \texttt{NVIDIA A800 80GB} GPUs. In this paper, we use \texttt{BM25} as the retriever and employ \texttt{LLaMA-3-8B} as the base model for all methods. Given that documents in the WikiWeb-ERP dataset are inherently concise, with lengths consistently below the standard chunk size of 10,000 tokens, we treat each entire document as a single passage without further segmentation. In addition, we dynamically set the thread count as the smaller value between the available cores and the number of independent nodes in the current layer, typically ranging from 4 to 7, which helps avoid thread-switching overhead while maximizing concurrency. Throughout our pipeline, we use \texttt{GPT-4o-mini} for atomic query generation, relation preprocessing, context-aware merging, and query rewriting. We set the scaling factor $alpha$ in Equation~\ref{eq2} to 10 through grid search to avoid deviating semantically.

For ChainRAG, DualRAG, KiRAG and LEGO-GraphRAG, we reproduce the results on our dataset WikiWeb-ERP using the standard implementations released by the original authors. For ChainRAG, we adopt the context integration setting that utilizes contexts retrieved by sub-questions to generate answers~\cite{zhu-etal-2025-mitigating}. For RetGen, GenGround and HopRAG, we replicate the experiments on our dataset following the parameter settings reported in their papers. For LEGO-GraphRAG, we build graphs from the corpus via Microsoft GraphRAG~\cite{edge2025localglobalgraphrag} before retrieval.

\subsection{Main Results}
As shown in Table~\ref{tab3}, our approach (\emph{PlanRAG w/ Ret}) achieves state-of-the-art performance on our WikiWeb-ERP dataset\footnote{Given that existing RAG datasets contain a relatively low proportion of ERPs, we conduct experiments on our dataset WikiWeb-ERP. This choice is deliberate, as conventional benchmarks are not well suited for evaluating methods tailored for ERPs.}, with the best results highlighted in bold and the second-best results underlined for clarity, where w/ Ret and w/o Ret indicate external retrieval is enabled and absent in PlanRAG framework\footnote{In the subsequent introduction, ``PlanRAG'' refers to ``PlanRAG w/ Ret''.}, respectively. The results show that compared with DirectLLM and NaiveRAG, PlanRAG w/ Ret achieves a significant gain, demonstrating the effectiveness of our framework on ERPs. PlanRAG w/ Ret also outperforms several advanced iteration-based RAG methods, including RetGen, GenGround, DualRAG, and KiRAG, which generally follow a chain-like process of decomposition, retrieval and generation. However, such iterative mechanisms often lack global planning between retrieval and generation, resulting in irrelevant document retrieval and error accumulation. In contrast, our framework performs global planning before generation, optimizing LQTs through dynamic programming and ensuring better alignment between retrieval and generation. Moreover, our framework surpasses graph-based RAG methods (ChainRAG, LEGO-GraphRAG and HopRAG), which leverage graph structures to retrieve logically connected information for multi-hop QA.  While these methods improve retrieval recall by exploring graph neighborhoods, they often introduce computational overhead for graph construction and traversal, and may still propagate errors through spurious connections. The global planning mechanism in PlanRAG directly optimizes the retrieval path through dynamic programming, thus leading to more robust performance. Finally, the fact that PlanRAG w/ Ret and PlanRAG w/o Ret both outperform NaiveRAG and RetGen, confirming that \textbf{the performance improvement is attributed to our planning mechanism rather than retrieval alone}.

\begin{table}[t]
\centering
\caption{Performance of the compared retrieval methods on our WikiWeb-ERP.}\label{tab3}
\vspace{-0.2cm}
\begin{tabular}{l|cccc}
\toprule
\textbf{Methods} & \textbf{Acc} & \textbf{F1} & \textbf{EM} & \textbf{Acc$\dagger$} \\
\midrule
DirectLLM & 12.31 & 16.55 & 10.48 & 16.88 \\
NaiveRAG & 17.54 & 20.33 & 14.65 & 22.19 \\
\midrule
\rowcolor[rgb]{ .9,  .9,  .9}
\multicolumn{5}{c}{\textit{Iteration-based methods}}\\
RetGen &19.34 &23.45 &15.78 &24.76 \\
GenGround &21.97 &23.21 &18.44 &27.54 \\
KiRAG & 24.72 & 27.83 & 19.65 & 27.92 \\
DualRAG &24.85 &26.65 &20.47 &\underline{28.35} \\
\midrule
\rowcolor[rgb]{ .9,  .9,  .9} 
\multicolumn{5}{c}{\textit{Graph-based methods}}\\
HopRAG&22.43  &24.97  &20.43  &27.14 \\
ChainRAG&24.34 &25.56  &20.23  &26.14 \\
LEGO-GraphRAG&\underline{25.48} &\underline{28.92}  &\underline{21.27}  &28.30 \\
\midrule
PlanRAG w/o Ret & 21.28 & 25.95 & 19.82 & 26.74\\
PlanRAG w/ Ret & \textbf{26.43}  &\textbf{30.15}  &\textbf{23.36}  &\textbf{30.68} \\
\bottomrule
\end{tabular}
\vspace{-0.2cm}
\end{table}

\begin{table}[t]
\centering
\caption{Ablation analysis of LQT construction, where rp, cp and ca denote relation preprocessing, cycle prevention and context-aware merging, respectively. In addition, ts, sd, td, tb and ss denote tree size, structural density, tree depth, tree balance and semantic similarities, respectively. }
\label{tab4}
\vspace{-0.2cm}
\begin{tabular}{l|cccc}
\toprule
\textbf{Methods} & \textbf{Acc} & \textbf{F1} & \textbf{EM} & \textbf{Acc$\dagger$} \\
\midrule
PlanRAG & \textbf{25.43} & \textbf{27.15} & 20.36 & \textbf{29.68} \\ \midrule
\quad w/o rp & 25.32 & 27.02 & \textbf{20.47} & 29.49 \\
\quad w/o cp & 22.32 & 23.45 & 17.77 & 25.19 \\
\quad w/o ca & 23.64 & 24.60 & 18.88 & 23.87 \\
\midrule
\quad w/o ts & 23.85 & 25.66 & 18.94 & 27.91 \\
\quad w/o sd & 24.12 & 25.89 & 19.27 & 28.35 \\
\quad w/o td & 23.97 & 25.73 & 19.05 & 28.12 \\
\quad w/o tb & 24.28 & 26.04 & 19.41 & 28.57 \\
\quad w/o ss & 21.64 & 23.31 & 16.89 & 25.43 \\
\bottomrule
\end{tabular}
\vspace{-0.2cm}
\end{table}

\subsection{Ablation Study}
We perform ablation studies on LQT construction to verify the effectiveness of our proposed method. As shown in Table~\ref{tab4}, we remove relation preprocessing, cycle prevention and context-aware merging to evaluate their impacts on the overall performance of PlanRAG. Notably, removing relation preprocessing (w/o pc) does not cause a significant performance degradation, because relation preprocessing primarily reduces system cost (such as runtime, token cost and the number of LLM calls) rather than directly improving generation quality. Removing cycle prevention (w/o cp) causes significant performance degradation, demonstrating that the mechanism is critical for maintaining LQT structure and preventing logical errors induced by cycles. Similarly, disabling context-aware merging (w/o ca) leads to a performance drop, highlighting the importance of the partially constructed tree to maintain semantic coherence throughout the planning process.

To evaluate the effectiveness of the cost model in guiding LQT construction, we also evaluate the performance of PlanRAG's ablated variants without each dimension in the cost model. Across all variants, the results in Table~\ref{tab4} consistently show that removing any single dimension leads to a degradation in overall performance, indicating that each contributes to the optimization objective. These findings highlight the significance of incorporating all five dimensions to achieve structural robustness and semantic soundness in LQT construction. Notably, the exclusion of the semantic similarity (w/o ss) results in the most significant decline. While structural regularization terms (e.g., tree depth or balance) in the cost model regulate execution efficiency, semantic similarity serves as a global alignment signal that prevents LQTs from drifting away from the original intent, suggesting that \textbf{effective planning for ERPs requires semantic anchoring}.

\begin{table}[t]
\centering
\small
\caption{Performance comparisons across different base models.}\label{tab5}
\begin{tabular}{l |l| c c c c}
\toprule
\textbf{Base Models} & \textbf{Methods} & \textbf{Acc} & \textbf{F1} & \textbf{EM} & \textbf{Acc$\dagger$} \\
\midrule
\multirow{3}{*}{Llama-3-70B} & NaiveRAG & 21.87 & 24.26 & 18.43 & 26.72 \\
& PlanRAG w/o Ret &25.34 & 28.32 & 24.08 & 29.81 \\
& PlanRAG w/ Ret & \textbf{31.64} & \textbf{33.28} & \textbf{26.17} & \textbf{35.55} \\ \midrule
\multirow{3}{*}{Llama-2-7B} & NaiveRAG & 15.32 & 18.67 & 12.91 & 20.14 \\
& PlanRAG w/o Ret &20.56 & 22.54 & 17.15 & 25.51\\
& PlanRAG w/ Ret & \textbf{24.25} & \textbf{26.48} & \textbf{20.03} & \textbf{28.39} \\ \midrule
\multirow{3}{*}{Llama-2-13B} & NaiveRAG & 16.95 & 19.84 & 14.27 & 21.83 \\
& PlanRAG w/o Ret&22.34 & 25.56 & 18.97 & 26.32\\
& PlanRAG w Ret& \textbf{26.61} & \textbf{27.83} & \textbf{21.82} & \textbf{29.67} \\ \midrule
\multirow{3}{*}{Llama-2-70B} & NaiveRAG & 20.38 & 23.45 & 17.68 & 25.52 \\
& PlanRAG w/o Ret  &24.85  &28.10  &21.05  &29.15\\
& PlanRAG w/ Ret& \textbf{30.43} & \textbf{31.97} & \textbf{25.35} & \textbf{35.94} \\ \midrule
\multirow{3}{*}{Qwen-2.5-7B} & NaiveRAG & 16.73 & 19.38 & 14.05 & 21.42 \\
& PlanRAG w/o Ret &21.18 &24.35 &18.02 &26.10\\
& PlanRAG w/ Ret& \textbf{25.96} & \textbf{27.74} & \textbf{21.28} & \textbf{29.85} \\ \midrule
\multirow{3}{*}{Qwen-2.5-14B} & NaiveRAG & 18.46 & 21.27 & 15.79 & 23.33 \\
& PlanRAG w/o Ret &24.85 &25.78 &20.64 &27.10\\
& PlanRAG w/ Ret& \textbf{28.28} & \textbf{30.13} & \textbf{24.45} & \textbf{32.62} \\ \midrule
\multirow{3}{*}{Qwen-2.5-72B} & NaiveRAG & 22.15 & 24.83 & 19.02 & 27.34 \\
& PlanRAG w/o Ret &27.70 &30.25 &23.35 &31.19\\
& PlanRAG w/ Ret & \textbf{32.72} & \textbf{34.19} & \textbf{26.96} & \textbf{37.23} \\
\bottomrule
\end{tabular}
\end{table}

\subsection{Robustness Across Different Base Models and Retrievers}
\paragraph{Base Models}
To verify the generalization ability of our framework across different LLMs, we also conduct experiments using various LLMs as base models, including Llama-3-70B, Llama-2 (7B/13B/70B), and Qwen-2.5 (7B/14B/72B). For each base model, we compare the performance of NaiveRAG, PlanRAG w/o Ret and PlanRAG w/ Ret. As shown in Table~\ref{tab5}, PlanRAG w/ Ret consistently yields performance improvements across all model scales and base models, demonstrating that \textbf{the effectiveness of our framework is model-agnostic}. Notably, the improvements are more significant for the models at smaller parameter scales (e.g., 7B and 13B), suggesting that our planning strategy compensates for their limited parametric knowledge and capacity. This observation highlights the potential of our method to enhance RAG systems even under constrained computational budgets.

\begin{table}[t]
\centering
\caption{Performance comparisons across different retrievers.}\label{tab6}
\begin{tabular}{l |l| c c c c}
\toprule
\textbf{Retrievers} & \textbf{Methods} & \textbf{Acc} & \textbf{F1} & \textbf{EM} & \textbf{Acc$\dagger$} \\
\midrule
\multirow{3}{*}{BGE-small} & NaiveRAG & 16.32 & 18.95 & 13.42 & 20.87 \\
& PlanRAG w/o Ret &21.40 &23.05 &17.60 &26.85\\
& PlanRAG w/ Ret & \textbf{25.15} & \textbf{26.28} & \textbf{20.74} & \textbf{29.43} \\ \midrule
\multirow{3}{*}{Contriever}& NaiveRAG & 21.83 & 24.36 & 18.75 & 27.91 \\
& PlanRAG w/o Ret &24.85 &26.85 &20.35 &29.77\\
& PlanRAG w/ Ret & \textbf{27.62} & \textbf{29.34} & \textbf{22.18} & \textbf{31.85} \\ \midrule
\multirow{3}{*}{ColBERT} & NaiveRAG & 22.17 & 24.86 & 19.32 & 27.45 \\
& PlanRAG w/o Ret & 27.48 & 29.47 & 23.14 & 31.80\\
& PlanRAG w/ Ret & \textbf{31.25} & \textbf{33.07} & \textbf{26.63} & \textbf{35.14} \\
\bottomrule
\end{tabular}
\end{table}

\paragraph{Retrievers} 
In addition to varying base models, we further evaluate our framework across diverse retrievers, including BGE-small~\cite{10.1145/3626772.3657878}, Contriever~\cite{izacard2021unsupervised}, and ColBERT~\cite{10.1145/3397271.3401075}. The experimental results in Table~\ref{tab6} consistently indicate that our framework surpasses NaiveRAG and \emph{PlanRAG w/o Ret} across all retrievers, demonstrating that \textbf{our framework's performance is robust and retriever-agnostic}.

\begin{table}[t]
\centering
\caption{Efficiency gains from relation preprocessing and parallel execution in PlanRAG, respectively, where rp denotes relation preprocessing.}
\label{tab7}
\small
\begin{tabular}{lccc}
\toprule
\textbf{Settings} & \textbf{Time (s)} & \textbf{Token Cost} & \textbf{LLM Calls} \\
\midrule
\rowcolor[rgb]{ .9,  .9,  .9}
\textit{Relation Preprocessing} & & & \\
\quad w/o rp & 82.8 & 3,454.4 & 63.6 \\
\quad w/ rp  & 8.4  & 1,265.2 & 10.4 \\
\midrule
\rowcolor[rgb]{ .9,  .9,  .9}
\textit{Parallel Execution} & & & \\
\quad w/o parallel & 18.7 & 2734.2 & 23.6 \\ 
\quad w/ parallel  & 7.3 & 2698.6 & 24.2 \\
\bottomrule
\end{tabular}
\end{table}

\subsection{Efficiency Analysis}
\paragraph{Relation Preprocessing's Impact on LQT Construction's Efficiency}
To verify the effect of relation preprocessing on LQT construction, we compare time, token cost and the number of LLM calls with and without the component. Table~\ref{tab7} reports the results on 100 sampled queries from WikiWeb-ERP. Obviously, enabling relation preprocessing reduces the average LQT construction time from 82.8s to 8.4s, decreases the number of LLM calls from 63.6 to 10.4, and lowers token cost from 3,454.4 to 1,265.2. Together with the results in Table~\ref{tab4}, \textbf{this optimization significantly reduces computational cost without sacrificing generation quality}.

\paragraph{Parallel Execution's Impact on LQT Execution's Efficiency}
To verify the effect of parallel execution on LQT execution, we compare single-threaded sequential execution (w/o parallel) with multi-threaded parallel execution (w/ parallel) for LQT execution on 100 randomly sampled queries from WikiWeb-ERP. Table~\ref{tab7} shows that parallel execution reduces the average execution time from 18.7s to 7.3s, achieving a 2.56$\times$ speedup. This improvement arises from the structural properties of LQTs, allowing nodes to be executed in parallel. Notably, the parallel execution primarily reduces execution time, while it does not reduce the number of LLM calls or token cost. Overall, these findings demonstrate that parallel execution is effective in enhancing efficiency, enabling PlanRAG to \textbf{achieve high accuracy while meeting real-time performance demands in practical deployments}.

\begin{figure*}
  \centering
  \includegraphics[width=\linewidth]{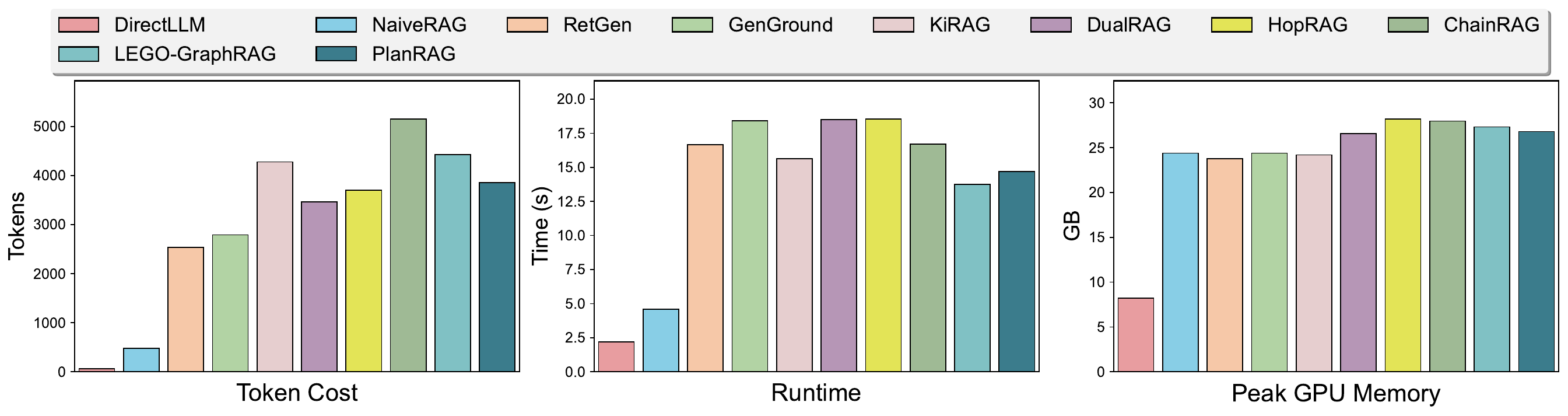}
  \caption{Comparisons of token cost, runtime, and GPU memory across various RAG methods (better viewed in color).}\label{figure3}
\end{figure*}

\paragraph{System Cost Comparisons}
To better evaluate the efficiency of PlanRAG, we compare its system cost with various baselines, reporting the average token cost, runtime, and peak GPU memory usage over 100 samples, respectively. Figure~\ref{figure3} shows clear divergences in system cost across different RAG systems. These iteration-based baselines rely on multiple rounds of retrieval and generation, where accumulated errors often trigger additional LLM calls, increasing token consumption. Their inherently sequential workflows prevent parallel execution, further increasing overall runtime. Graph-based baselines introduce substantial overhead in graph construction and traversal. As the graph grows larger, redundant nodes and irrelevant edges accumulate quickly, thereby inflating resource usage. Compared with these baselines, PlanRAG employs the global planning mechanism to reduce redundant retrieval and unnecessary inference. It makes the execution more focused and amenable to parallelization. Although planning introduces an initial cost, the overall workflow becomes more streamlined, \textbf{resulting in a superior cost–efficiency trade-off in terms of system cost and performance}.

\begin{figure}[t]
  \centering
  \includegraphics[width=\linewidth]{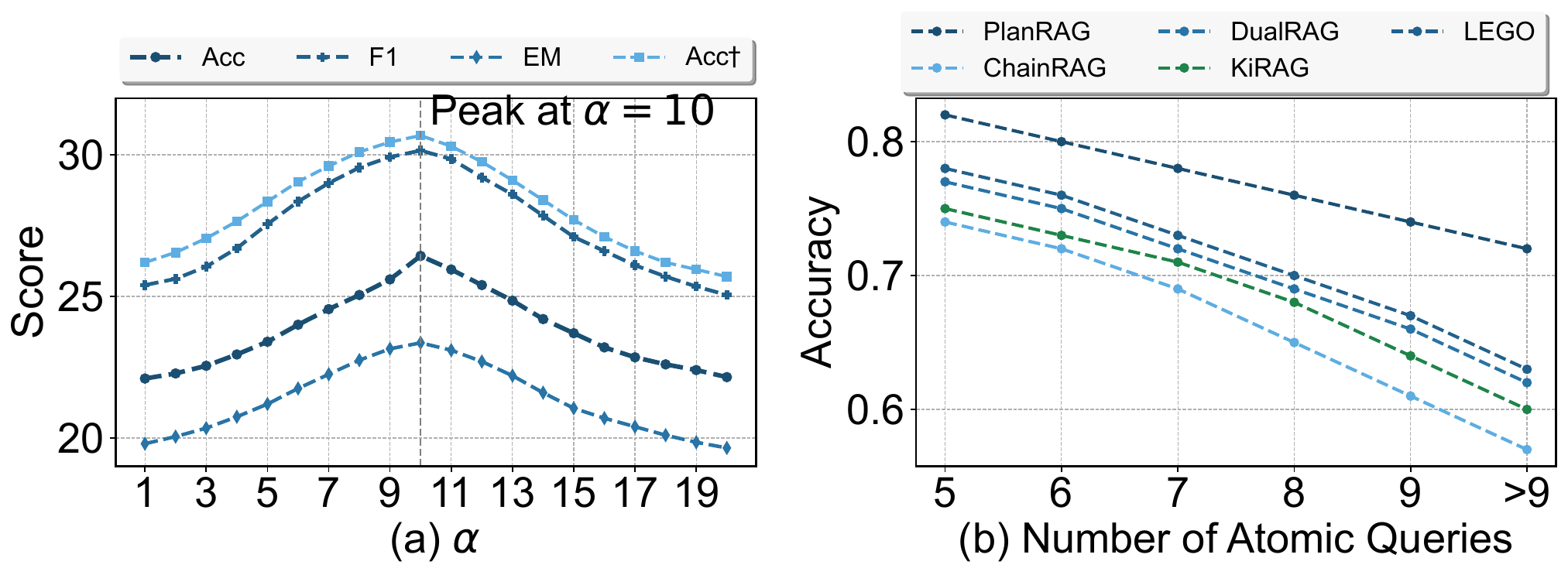}
  \vspace{-0.2cm}
  \caption{(a) Performance of PlanRAG across different scaling factor $\alpha$. (b) Performance of various RAG methods across different query complexities.}\label{figure4}
\end{figure}

\subsection{Results on Multi-Hop QA}
\begin{table}[t]
\centering
\caption{Performance comparison (Acc) of the compared retrieval methods on four Multi-hop QA datasets. Hot, MuSi, 2Wiki, Strategy denote HotpotQA, MuSiQue, 2WikiMultihopQA, and StrategyQA, respectively.}\label{tab8}
\vspace{-0.2cm}
\begin{tabular}{l|cccc}
\toprule
\textbf{Methods} & \textbf{Hot} & \textbf{MuSi} & \textbf{2Wiki} & \textbf{Strategy} \\
\midrule
DirectLLM & 32.47 & 8.73 & 30.18 & 62.41 \\
NaiveRAG  & 38.92 & 10.54 & 36.83 & 66.68 \\
\midrule
\rowcolor[rgb]{ .9,  .9,  .9}
\multicolumn{5}{c}{\textit{Iteration-based methods}}\\
RetGen    & 42.13 & 11.79 & 39.46 & 68.94 \\
GenGround & 44.21 & 12.18 & 41.29 & 70.31 \\
KiRAG     & 46.17 & \underline{15.33} & 43.47 & 72.08 \\
DualRAG   & \textbf{48.52} & 14.87 & 45.23 & 73.38 \\
\midrule
\rowcolor[rgb]{ .9,  .9,  .9} 
\multicolumn{5}{c}{\textit{Graph-based methods}}\\
HopRAG        & 46.79 & 14.63 & 45.88 & 73.02 \\
LEGO-GraphRAG    & 47.58 & 15.21 & \underline{47.09} & 74.17 \\
ChainRAG & 48.43 & \textbf{15.83} & \textbf{49.07} & \underline{75.04} \\
\midrule
PlanRAG & \underline{47.91} & 15.12 & 45.82 & \textbf{76.27} \\
\bottomrule
\end{tabular}
\vspace{-0.2cm}
\end{table}
To further validate the generality of PlanRAG beyond ERPs, we evaluate our method on four widely used multi-hop QA benchmarks, including HotpotQA~\cite{yang-etal-2018-hotpotqa}, MuSiQue~\cite{trivedi2022musique}, 2WikiMultiHopQA~\cite{ho-etal-2020-constructing}, and StrategyQA~\cite{geva2021did}. As shown in Table~\ref{tab8}, PlanRAG achieves competitive performance across most datasets and obtains the best result on StrategyQA, demonstrating its generalization ability beyond ERPs. In particular, PlanRAG performs comparably to graph-based baselines on HotpotQA and MuSiQue, but shows a gap on 2WikiMultiHopQA, whose reasoning chains are more explicit and therefore better suited to graph-based retrieval methods. In contrast, PlanRAG is mainly designed for ERPs with ambiguous dependencies, where explicit reasoning chains are not present, and our method can perform explicit query planning to effectively improve performance.
\section{Further Analysis and Discussions}
\subsection{Sensitivity Analysis of Scaling Factor $\alpha$}
To further validate the effectiveness of the scaling factor $\alpha$ in the cost model, we conduct a sensitivity analysis. Specifically, we vary $\alpha$ within the range $[1, 20]$ and evaluate the final accuracy. Figure~\ref{figure4}(a) shows that performance first improves and then stabilizes as $\alpha$ increases, reaching the highest accuracy at $\alpha = 10$; further increasing $\alpha$ yields limited gains or slight degradation. From a scaling perspective, the structural terms (tree depth or balance) grow with the number of nodes, whereas the semantic similarity term is normalized within $[0,1]$, leading to a scale mismatch. The introduction of $\alpha$ serves to align their contributions. These results indicate that $\alpha$ is not an arbitrary tuning parameter, but plays a critical role in balancing structural regularization and semantic alignment, with its optimal value corresponding to a well-calibrated trade-off between the two.
\subsection{Performance and Cost Comparisons Across Query Complexity}
\paragraph{Performance Comparisons Across Query Complexity}
To further examine the robustness of PlanRAG varying query complexity\footnote{We define query complexity as the number of atomic queries decomposed from ERPs.}, we categorize queries into the buckets of size $\{5, 6, 7, 8, 9, >9\}$ based on the number of atomic queries, and report the average accuracy within each bucket. This setup reflects the intuition that a larger number of atomic queries generally implies deeper dependency chains. The results in Figure~\ref{figure4}(b) show that PlanRAG consistently outperforms the baselines across all buckets. Notably, while all methods exhibit performance degradation as the number of atomic queries increases, PlanRAG demonstrates a significantly slower decline. PlanRAG explicitly encodes dependencies and performs global structural optimization, which localizes errors and prunes suboptimal plans early. Moreover, its structural regularization terms (e.g., tree depth and balance) in the cost model prevent planning from degenerating into long chains, leading to more stable performance under high complexity. Overall, these results emphasize the effectiveness of PlanRAG in scaling to high-complexity queries beyond conventional RAG pipelines.

\begin{figure}[htpb]
  \centering
  \includegraphics[width=\linewidth]{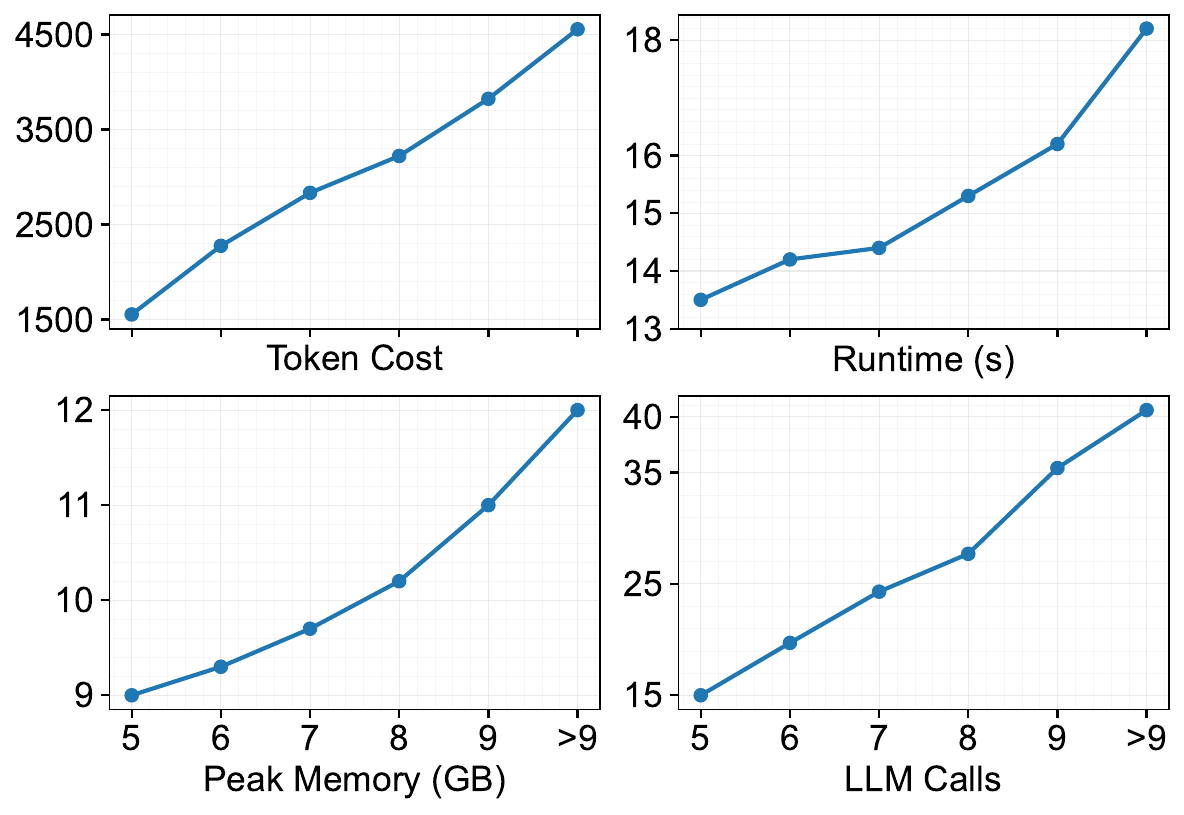}
  \vspace{-0.2cm}
  \caption{System cost of PlanRAG when varying query complexity. 
  }\label{figure5}
\end{figure}

\paragraph{System Cost Comparisons Across Query Complexity} 
To assess the system cost of PlanRAG when increasing query complexity, we analyze system cost as a function of the number of atomic queries $n$. We bucket the ERPs in WikiWeb-ERP according to $n \in \{5, 6, 7, 8, 9, >9\}$ and measure four metrics: (a) token cost, (b) runtime, (c) peak GPU memory, and (d) the number of LLM calls. As shown in Figure~\ref{figure5}, the values of all four metrics increase as the number of atomic queries $n$ grows. Specifically, token cost and the number of LLM calls grow approximately linearly with $n$, reflecting the expansion of intermediate steps as LQTs increase in size. Runtime also increases with $n$, while peak GPU memory rises from around 9.0~GB at $n=5$ to nearly 12~GB at $n>9$, due to the accumulation of KV cache and intermediate results during execution. Overall these results demonstrate that PlanRAG exhibits a predictable increase in system cost as query complexity grows, confirming its scalability for complex queries.

\begin{table}[t]
\centering
\caption{Human and LLM (GPT-4o) evaluation scores for the LQTs generated by our method and by LLM, respectively.}
\vspace{-0.2cm}
\label{tab9}
\begin{tabular}{lcc}
\toprule
\textbf{Methods} & \textbf{Human} & \textbf{LLM (GPT-4o)} \\
\midrule
LLM Generation & 3.22 & 3.17 \\
Our Method w/o ss & 3.26 & 3.37 \\
Our Method  & \textbf{3.52} & \textbf{3.54} \\
\bottomrule
\end{tabular}
\vspace{-0.2cm}
\end{table}

\subsection{Evaluation of LQT Quality}
To examine the effectiveness of our LQT construction method, we conduct a comparative evaluation between LQTs generated by our method and those directly produced by LLMs. The evaluation prompts for LLMs are listed in Appendix A, which measures whether LQTs exhibit structural robustness and semantic soundness. Each LQT's quality is evaluated using a 4-point ($1\sim 4$ score) ordinal scale as below, by human annotators and the LLM judge, respectively.

\noindent\textbf{4 – Excellent:} The LQT exhibits a well-formed and stable structure with clearly defined dependencies among atomic queries, while maintaining strong semantic alignment with the original query intent throughout the tree.

\noindent\textbf{3 – Good:} The LQT is largely structurally coherent and semantically consistent, with minor structural imperfections or slight semantic ambiguities that do not undermine the overall process.

\noindent\textbf{2 – Fair:} The LQT shows noticeable structural weaknesses (e.g., missing or improper dependencies) or semantic inconsistencies that partially affect the correctness of the workflow.

\noindent\textbf{1 – Poor:} The LQT lacks structural robustness and semantic soundness, exhibiting disorganized dependencies and significant semantic drift.

The final score $S$ is computed as the average score over all 30 generated LQTs: \( S = \frac{1}{30} \sum_{i=1}^{30} s_{i}. \) where $s_i$ denotes the score assigned to the LQT. Table~\ref{tab9} reports the scores of human assessment and LLM judge, respectively.  The results show that the LQTs constructed by our method have better quality than the LQTs directly generated by the LLM. Furthermore, we construct an ablated variant by removing the semantic similarity from the cost model and evaluate the resulting LQTs. The quality drops noticeably, indicating that semantic similarity plays a critical role in maintaining the semantic alignment of LQTs. Specifically, a powerful LLM could directly generate a plan in a single step. However, this approach often lacks stable structural constraints, and different invocations may produce inconsistent or dependency-violating plans. In contrast, we first decompose complex queries into atomic queries and then construct an LQT with explicit structural constraints, ensuring clear dependencies and executable reasoning.

The linear graph sometimes generated by LLMs is not necessarily optimal for ERPs. In many cases, some atomic queries are semantically independent and can be executed in parallel. For example, in Figure~\ref{figure1}, identifying candidate authors (and their death years) and retrieving candidate chronologies are largely independent tasks and can be performed separately before being aligned through the constraint that the chronology’s final year matches the author’s death year. Therefore, LQT organizes these queries into a tree rather than a strict linear chain, reducing sequential dependency and error propagation. Notably, the linear structure generated by Claude resembles the sequential reasoning strategy used in baselines such as ReTGen and GenGround, while PlanRAG significantly outperforms them in Table~\ref{tab3}, highlighting the benefit of tree-structured planning.

\subsection{Fine-grained Analysis on LQT Construction}\label{fine-grainedanalysisonlqtconstruction}
To evaluate LQT construction more comprehensively, we further analyze 100 randomly sampled queries from four critical aspects: atomic query generation, atomic query coverage, linear degradation rate, and relationship classification accuracy. The evaluation results are listed in Table~\ref{tab10}. \vspace{-0.25cm}
\paragraph{Atomic Query Generation} To evaluate the accuracy of atomic query generation, we conduct a manual assessment on 100 randomly sampled queries. The evaluation results indicate that the atomic queries generated by LLMs achieve 98.5\% accuracy. The reported 98.5\% accuracy is obtained through manual annotation of the generated atomic queries and their corresponding triples, where multiple annotators independently evaluate the correctness of each triple based on predefined criteria. Specifically, inter-annotator agreement is assessed using Fleiss's kappa, achieving $k$ = 0.82, which indicates strong agreement among annotators. It demonstrates that our decomposition method can extract the minimal semantic units from the original ERPs, thereby providing a reliable foundation for the subsequent LQT construction.\vspace{-0.25cm}
\paragraph{Atomic Query Coverage} To verify that LQTs have high atomic query coverage, we systematically count the number of atomic queries in all generated LQTs. The results show that PlanRAG achieves full coverage of all atomic queries in 94.2\% of cases. It demonstrates that our planning mechanism can integrate the vast majority of atomic queries, thereby supporting execution grounded in a complete chain of atomic queries. \vspace{-0.25cm}
\paragraph{Linear Degradation Rate} Given the ERPs characterized by inherent high uncertainty and difficulty in reduction, our planning mechanism explicitly organizes dependencies through LQT's structure. In the ideal case where semantic relationships between atomic queries are clear and easily distinguishable, PlanRAG constructs well-structured LQTs. However, in the worst case, where atomic queries are largely semantically independent, LQTs may degrade into an approximately linear chain structure. To quantify the frequency of such occurrences, we conduct structural analysis on the generated LQTs. The results show that only 3.2\% of queries exhibit a near-linear structure, while the vast majority of queries (over 96\%) maintain explicit tree-like dependency relationships. \vspace{-0.25cm}
\paragraph{Relationship Classification} As highlighted in Section~\ref{method}, we leverage LLMs to identify unrelated relationships in relation preprocessing. To assess the reliability of the classification, we manually verify the predicted semantic relationships and measure their classification accuracy. The results show that LLMs achieve 96.4\% classification accuracy for unrelated relationships, while the accuracies for parent-child and sibling relationships are 83.1\% and 87.9\%, respectively. High accuracy for unrelated relationships demonstrates that relation preprocessing can improve the efficiency of LQT construction.
\begin{table}[t]
\centering
\caption{Fine-grained analysis of LQT construction.}
\label{tab10}
\vspace{-0.2cm}
\begin{tabular}{lc}
\toprule
\textbf{Analysis Aspects} & \textbf{Results} \\
\midrule
Atomic Query Generation & 98.5\% \\
Atomic Query Coverage  & 94.2\% \\
Linear Degradation Rate& 3.2\%  \\
\midrule
\rowcolor[rgb]{ .9,  .9,  .9}
\textit{Relationship Classification} & \\
\quad Unrelated     & 96.4\% \\
\quad Parent--Child & 83.1\% \\
\quad Sibling       & 87.9\% \\
\bottomrule
\end{tabular}
\end{table}
\subsection{Stage-level Robustness Analysis}
\begin{table}[t]
\centering
\small
\setlength{\tabcolsep}{6pt}
\renewcommand{\arraystretch}{1.1}
\caption{Stage-level robustness analysis measured by Accuracy under controlled perturbations at different stages of the PlanRAG pipeline.}\vspace{-0.2cm}
\label{tab11}
\begin{tabular}{c|ccc}
\hline
\textbf{Noise Level} & \textbf{Atomic Query} & \textbf{Relation} & \textbf{Execution} \\
\hline
0\%  & 27.65 & 27.65 & 27.65 \\
10\% & 24.84 & 25.22 & 24.50 \\
20\% & 22.81& 23.07 & 21.77 \\
30\% & 19.37 & 20.65 & 18.69 \\
\hline
\end{tabular}
\end{table}
To analyze stage-level robustness, we conduct controlled perturbation experiments to examine how errors propagate through the PlanRAG pipeline in Table~\ref{tab11}. We randomly sample 100 queries from the dataset and inject noise at three stages, with noise levels of 10\%, 20\%, and 30\%. (1) \textbf{Atomic query generation}: we randomly remove atomic queries to simulate incomplete or noisy decomposition. (2) \textbf{Relation classification}: we randomly select edges and replace their relations with randomly chosen labels (parent–child, sibling, or unrelated), thereby perturbing the tree structure. (3) \textbf{LQT execution}: we replace retrieved contents with irrelevant text to simulate retrieval and reasoning errors. For each stage, we vary the noise level while keeping the other stages unchanged and evaluate the impact on final answer accuracy. Results show that performance degrades gradually as noise increases, indicating that errors do propagate across stages. However, the degradation is smooth rather than catastrophic, indicating that PlanRAG is robust to upstream errors, mainly due to its structured planning.
\subsection{Error Analysis}
To evaluate the credibility of PlanRAG on ERPs, we categorize the errors in queries into four types: ($i$) \textbf{Planning errors (36\%)} primarily stem from insufficient recognition of dependencies among atomic queries, resulting in LQTs that can not accurately capture grounding logical orders, thereby compromising overall performance. These errors are typically caused by ambiguous relations or low-confidence merging decisions, highlighting the necessity of more accurate relationship classification. ($ii$) \textbf{Retrieval errors (32\%)} arise when retrievers only acquire partially relevant but ultimately irrelevant information, or fail to retrieve critical evidence passages, reducing the support for subsequent steps. These errors are often caused by entity mismatches or inadequate query reformulation, suggesting that retrieval robustness can be improved through hybrid retrieval and optimized query rewriting. ($iii$) \textbf{Reasoning consistency errors (19\%)} occur when individual answers of atomic queries are correct, but their integration into a global reasoning outcome leads to deviations from factual conclusions, reflecting the limitations in global consistency. ($iv$) \textbf{Generation and synthesis errors (13\%)} are observed during the final answer aggregation stage, manifesting as ambiguous entity references or disrupted output formats, typically caused by the instability in structured representations over long reasoning chains.

Overall, our PlanRAG's main bottlenecks lie in planning and retrieval, and further performance improvements require the reinforcement of adaptive planning mechanisms and robust retrieval strategies.

\section{Conclusion}
In this paper, we propose PlanRAG, an end-to-end RAG framework designed for ERPs motivated by database query planning. PlanRAG first decomposes ERPs into atomic queries. These atomic queries are then organized into optimized LQTs that model semantic dependencies via dynamic programming, mitigating retrieval noise and error accumulation. Based on optimized LQTs, PlanRAG iteratively executes aggregation, rewriting, retrieval, and generation in a bottom-up and parallel manner, recursively propagating intermediate results to efficiently yield the final answer. Extensive experiments on the newly released WikiWeb-ERP dataset demonstrate consistent improvements over iteration-based and graph-based RAG baselines, while further analyses confirm the efficiency, effectiveness and interpretability of our proposed planning mechanism.\\
\indent We acknowledge that our method has several limitations. First, PlanRAG struggles to generalize to simple problem scenarios, as its planning mechanism may be unnecessary for straightforward queries. Second, for queries with low logical complexity or clear stepwise paths, the method may degenerate into a standard multi-hop QA decomposition paradigm. Additionally, our cost model is a simplified abstraction and does not explicitly normalize costs based on factors such as document length or task complexity. Its goal is not to precisely estimate the true execution cost of each semantic query, but rather to provide a lightweight heuristic to guide plan selection. While different subplans may share the same modeled cost yet differ in actual execution overhead, variations in retrieval and reasoning cost are partially reflected through semantic similarity and structural terms, which help prioritize more coherent and efficient plans. Moreover, we plan to incorporate fine-tuning strategies to reduce reliance on LLMs in components such as atomic query generation, relation preprocessing, and query rewriting, by replacing GPT-4o-mini with lightweight open-source models (e.g., LLaMA and Qwen families). Lastly, we will explore how to map LQTs into a physical query plan, incorporating system-level optimizations such as caching mechanisms to provide a more complete systems-level optimization formulation.
\bibliographystyle{ACM-Reference-Format}
\bibliography{sample-base}
\appendix
\twocolumn
\section{Prompts}\label{prompts}
\begin{tcolorbox}[
    enhanced,
    colback=white,
    colframe=blue!50!black,
    title=Prompts for Atomic Query Generation,
    fonttitle=\bfseries,
    fontupper=\small,
    listing options={breaklines=true}  
]
Please decompose the following problem into independent, retrievable atomic queries $Q_{1}$…$Q_{n}$\\
Each atomic query must meet ALL requirements below:\\
1. Contain only one triple (subject, relation, object).\\
2. No pronouns or ambiguous references — replace with full entities from the problem.\\
3. Be directly retrievable from a KB or text corpus.\\
4. Describe clearly what to retrieve, in natural language.\\
5. Preserve ALL information in the original problem — do not omit any entity or relation.\\
6. Output must be in strict JSON format as shown in the examples.\\
-----\\
$[$[Examples]$]$\\
-----\\
Problem: [[Problem]]\\
Atomic Queries:\\
\end{tcolorbox}
\begin{tcolorbox}[
    enhanced,
    colback=white,
    colframe=blue!50!black,
    title=Prompts for Determining Relationships between Atomic Queries,
    fonttitle=\bfseries,
    fontupper=\small,
    listing options={breaklines=true}  
]
For the given problem: [[query]]\\
After decomposition, the atomic queries are: [[atomic queries]]\\
Please determine semantic relationships type between two atomic queries:\\
1 (Parent-Child Relationship): One query is the child node of the other (similar to a logical plan tree parent-child relationship).\\
In this case, you must also output direct = 3 or 4:\\
    3 means Atom1 is the child of Atom2.\\
    4 means Atom2 is the child of Atom1.\\
2 (Sibling Relationship): The two queries are sibling nodes, describing the same subject or parallel conditions.\\
0 (Unrelated Relationship): The two queries describe completely different content and have no semantic connection.\\
The output format must be strictly JSON, for example: 
\verb|{"relation_type": 1, "direct": 3}| or \verb|{"relation_type": 2}| or \verb|{"relation_type": 0}|\\
-----\\
$[$[Examples]$]$\\
-----\\
The currently constructed logical query tree is: \textbf{[[built\_tree]]}\\
Determine the relationship between the following two atomic queries:\\
Atom Query 1: [[atom1]]\\
Atom Query 2: [[atom2]]\\
\end{tcolorbox}

\begin{tcolorbox}[
    enhanced,
    colback=white,
    colframe=blue!50!black,
    title=Prompts for Converting Logical Query Trees into Natural Language Queries,
    fonttitle=\bfseries,
    fontupper=\small,
    listing options={breaklines=true}  
]
Convert the following logical query tree into a single coherent natural language query.\\
Requirements:\\
1. Integrate all nodes into one fluent question, preserving all entities and constraints.\\
2. Use clear, unambiguous language, no pronouns or omissions.\\
3. Output only the final natural language query — no explanation, no extra formatting.\\
-----\\
$[$[Examples]$]$\\
-----\\
Logical query tree: [[LogicalQueryTree]]\\
Natural language query:\\
\end{tcolorbox}

\begin{tcolorbox}[
    enhanced,
    colback=white,
    colframe=blue!50!black,
    title=Prompts for Rewriting Queries,
    fonttitle=\bfseries,
    fontupper=\small,
    listing options={breaklines=true}  
]
For the given original question: [[Query]]\\
The current decomposed atomic query is: [[automic\_query]]\\
You are also given the previous questions and their answers that led to this atomic query: [[previous\_his]]\\
Task: Rewrite the current atomic query by integrating relevant information from the previous questions and answers, so that the rewritten query becomes self-contained, specific, and contextually grounded.\\
-----\\
$[$[Examples]$]$\\
-----\\
Output format: \verb|{"answer": "<rewritten\_query>"}|\\
where \verb|<rewritten_query>| is the fully self-contained, contextually enriched query.\\
\end{tcolorbox}
\begin{tcolorbox}[
    enhanced,
    colback=white,
    colframe=blue!50!black,
    title=Prompts for Generating LQTs by LLMs,
    fonttitle=\bfseries,
    fontupper=\small,
    listing options={breaklines=true}  
]
You are given a natural language query that involves complex, multi-step reasoning.\\
Task:\\
Directly construct a Logical Query Tree (LQT) for the given query.\\
Definition of LQT:\\
- An LQT is a directed acyclic graph (DAG).\\
- Each node represents an atomic query corresponding to a single relational triple \\
  (subject, relation, object).\\
- Each edge represents a semantic dependency, where a child node provides necessary information for its parent.\\
-----\\
Examples:\\
Query: There was an early Christian poetic hymn composed by a late antique writer who passed away around the mid-5th century. The year of this writer’s death coincides with 
the last year of a scientific chronology that reconstructs environmental conditions from several centuries before the modern era. What is the name of this chronology?\\
LQT: \\
\begin{verbatim}{
    ''nodes``: [
        {''id``: ''Q1``, ''query``: ''What were the 
    early Christian hymns?``},
        {''id``: ''Q2``, ''query``: ''Which late   
    antique authors composed Christian hymns?``},
        {"id": ''Q3``, ''query``: ''Which authors  
    passed away around the middle of the 5th century?``},
        {"id": ''Q4``, ''query``: ''Which authors’ 
    years of death coincide with the final year 
                                    of a scientific chronology?``},
        {"id": ''Q5``, ''query``: ''Which     
    scientific chronologies reconstruct environmental 
    conditions from several centuries 
    before the modern era?``},
        {"id": ''Q6``, ''query``: ''What is the     
    name of the scientific chronology?``}
    ],
    ''edges``: [
        {''from``: ''Q1``, ''to``: ''Q2``},
        {''from``: ''Q2``, ''to``: ''Q4``},
        {''from``: ''Q3``, ''to``: ''Q4``},
        {''from``: ''Q4``, ''to``: ''Q6``},
        {''from``: ''Q5``, ''to``: ''Q6``}
    ] 
}
\end{verbatim}
-----\\
Query: [[Query]]\\
LQT: [[LQT]]\\
\end{tcolorbox}
\section{Algorithm}
\begin{algorithm}[htbp]
\caption{Logical Query Tree Construction}
\label{alg:lqt}
\begin{algorithmic}[1]
\Require Atomic queries $\mathcal{Q} = \{q_1, q_2, \ldots, q_n\}$
\Ensure Optimized LQT $T^\star$

\State \textbf{Relation Preprocessing}
\ForAll{$(q_i, q_j)$ with $i < j$}
    \State $\mathcal{R}_{i,j} \leftarrow \textsc{SemanticRelation}(q_i, q_j)$
\EndFor

\State \textbf{Initialization}
\ForAll{$q_i \in \mathcal{Q}$}
    \State $\textsc{DP}[\{q_i\}] \leftarrow \textsc{Tree}(q_i)$
    \State $\textsc{Cost}[\{q_i\}] \leftarrow 0$
\EndFor

\State \textbf{Dynamic Programming}
\For{$k = 2$ to $|\mathcal{Q}|$}
    \ForAll{subsets $S \subseteq \mathcal{Q}$ with $|S| = k$}
        \State $\textsc{Cost}[S] \leftarrow \infty$
        \ForAll{partitions $S = A \cup B$, $A \cap B = \emptyset$}
            \If{\textsc{ViolatesRelation}$(A, B, \mathcal{R})$}
                \State \textbf{continue}
            \EndIf
            \If{\textsc{CreatesCycle}$(\textsc{DP}[A], \textsc{DP}[B])$}
                \State \textbf{continue}
            \EndIf
            \State $\mathcal{T} \leftarrow \textsc{Merge}(\textsc{DP}[A], \textsc{DP}[B])$
            \State $C \leftarrow \textsc{Cost}[A] + \textsc{Cost}[B] + \textsc{TreeCost}(\mathcal{T})$
            \If{$C < \textsc{Cost}[S]$}
                \State $\textsc{Cost}[S] \leftarrow C$
                \State $\textsc{DP}[S] \leftarrow \mathcal{T}$
            \EndIf
        \EndFor
    \EndFor
\EndFor

\State \Return $\mathcal{T}^\star \leftarrow \textsc{DP}[\mathcal{Q}]$
\end{algorithmic}
\end{algorithm}
\section{Case Study}
\begin{figure*}[htbp]
  \centering
  \includegraphics[width=\linewidth]{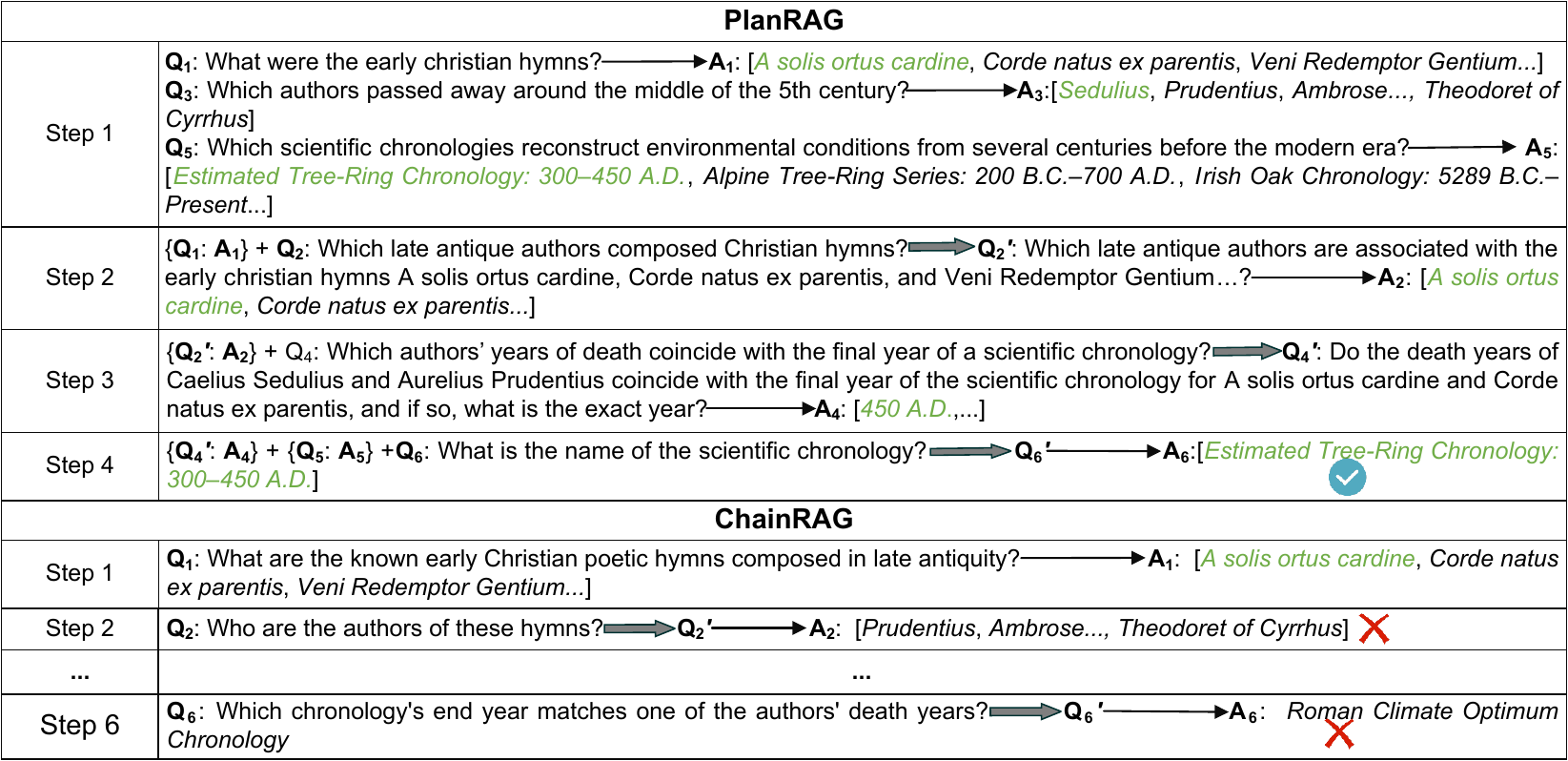}
  \caption{LQT execution comparison of PlanRAG and ChainRAG on an representative ERP introduced in Section 1. 
  \includegraphics[height=0.5em]{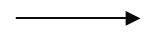} and \includegraphics[height=0.5em]{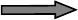} denote the generation and aggregation operation, respectively. For clarity, the retrieval operations are omitted. } \label{figure6}
\end{figure*}

To further demonstrate the advantages of our planning mechanism, we compare the detailed results of PlanRAG and ChainRAG on a toy example of ERP. As shown in Figure~\ref{figure6}\footnote{Please note that PlanRAG and ChainRAG generate different sets of atomic queries in the example.}, PlanRAG explicitly identifies dependencies among atomic queries and executes them in topological order, thereby preventing error propagation and ultimately producing the correct answer. In contrast, ChainRAG executes sub-queries sequentially without explicit dependency planning, introducing early errors to propagate and affecting the final result\footnote{Due to space limitations, the full execution flow of ChainRAG is not depicted in Figure~\ref{figure6}.}. This comparison result clearly demonstrates that \textbf{PlanRAG suppresses error propagation through explicit logical planning}, enhancing the accuracy and interpretability of RAG systems.

\end{document}